\documentclass[a4paper,fleqn,usenatbib,useAMS]{mnras}
\usepackage{graphicx}	
\usepackage{amsmath}	
\usepackage{amssymb}	
\usepackage{multicol}        
\usepackage{multirow}
\usepackage{pdflscape}	
\usepackage[T1]{fontenc}
\usepackage{ae,aecompl}
\usepackage{mathptmx}
\usepackage{natbib}
\usepackage{color}
\usepackage{xspace}
\usepackage{hyperref}
\usepackage{rotating}
\usepackage{dcolumn,longtable,booktabs}
\usepackage{lscape}
\usepackage{longtable}
\usepackage{supertabular}
\usepackage[flushleft]{threeparttable}
\usepackage[usenames,dvipsnames,svgnames,table]{xcolor}

\newcommand{\atlas}{ATLAS$^{\mathrm{3D}}$}
\newcommand{\hi}{{\sc H\,i}}
\title[SF associated with \hi \ in the outskirts of ETGs]{Star formation associated with neutral hydrogen in the outskirts of early-type galaxies}
\author[M.K. Y{\i}ld{\i}z et al.]{Mustafa K. Y{\i}ld{\i}z$^{1,2,3}$\thanks{E-mail: mkyildiz@astro.rug.nl}, Paolo Serra$^{3}$, Reynier F. Peletier$^{1}$, Tom A. Oosterloo$^{1,4}$, \and Pierre-Alain Duc$^{5}$\\\
$^{1}$Kapteyn Astronomical Institute, University of Groningen, PO Box 800, NL-9700 AV Groningen, the Netherlands\\
$^{2}$Astronomy and Space Sciences Department, Science Faculty, Erciyes University, Kayseri, 38039 Turkey\\
$^{3}$CSIRO Astronomy and Space Science, Australia Telescope National Facility, PO Box 76, Epping, NWS 1710, Australia\\
$^{4}$Netherlands Institute for Radio Astronomy (ASTRON), Postbus 2, NL-7990 AA Dwingeloo, the Netherlands\\
$^{5}$Laboratoire AIM Paris-Saclay, CEA/Irfu/SAp CNRS Universite Paris Diderot, F-91191 Gif-sur-Yvette Cedex, France\\}

\begin{document}

\date{Submitted 0 August 1234; Accepted 0 October 1234}
\maketitle

\label{firstpage}
\begin{abstract}
{About 20 percent of all nearby early-type galaxies ($M_{\star} \gtrsim~6~\times 10^{9}~\mathrm{M}_{\odot}$ ) outside the Virgo cluster are surrounded by a disc or ring of low-column-density neutral hydrogen (\hi) gas with typical radii of tens of kpc, much larger than the stellar body. In order to understand the impact of these gas reservoirs on the host galaxies, we analyse the distribution of star formation out to large radii as a function of \hi \ properties using GALEX UV and SDSS optical images. Our sample consists of 18 \hi-rich galaxies as well as 55 control galaxies where no \hi \ has been detected. In half of the \hi-rich galaxies the radial UV profile changes slope at the position of the \hi \ radial profile peak. To study the stellar populations, we calculate the FUV-NUV and UV-optical colours in two apertures, 1-3 and 3-10 $R_\mathrm{eff}$. We find that \hi-rich galaxies are on average 0.5 and 0.8 mag bluer than the \hi-poor ones, respectively. This indicates that a significant fraction of the UV emission traces recent star formation and is associated with the \hi \ gas. Using FUV emission as a proxy for star formation, we estimate the integrated star formation rate in the outer regions ($R > 1R_\mathrm{eff}$) to be on average $\sim$ 6$\times$10$^{-3}$ M$_{\odot}~\mathrm{yr}^{-1}$ for the \hi-rich galaxies. This rate is too low to build a substantial stellar disc and, therefore, change the morphology of the host. We find that the star formation efficiency and the gas depletion time are similar to those at the outskirts of spirals. 
}
\end{abstract}
\begin{keywords}
galaxies: elliptical and lenticular, cD -- galaxies: evolution -- galaxies: ISM -- galaxies: statistics: -- galaxies: stellar content
\end{keywords}

\section{Introduction}
\label{sec:introduction}

Since a long time evidence has appeared in the literature that early-type galaxies (ellipticals and lenticulars, hereafter ETGs) harbour central discs \citep{1951ApJ...113..413S, 1970ApJ...160..831S, 1976ApJ...206..883V}. In recent years, this has been quantified by several optical studies, based on broad band imaging and integral-field spectroscopy \citep[e.g.,][]{2011MNRAS.413..813C, 2011MNRAS.418.1452L, 2012ApJS..198....2K, 2013MNRAS.432.1768K,2014MNRAS.444.3340W}. Since they are relatively shallow and have a small field of view, these studies are limited to the inner regions of nearby ETGs. However, deep optical imaging, together with the kinematics of planetary nebulae and of globular clusters, have revealed that these discs can extend much further out \citep{2003Sci...301.1696R,2013MNRAS.428..389P,2014MNRAS.440.1458D,2015MNRAS.446..120D}.

Rotating discs are often detected in the neutral hydrogen (\hi) gas phase, too \citep[e.g.,][]{2006MNRAS.371..157M, 2007A&A...465..787O, 2010MNRAS.409..500O}. The most complete census of \hi \ discs in ETGs to date was obtained by \citet[][hereafter S12]{2012MNRAS.422.1835S} as part of the \atlas \ survey\footnote{http://www-astro.physics.ox.ac.uk/atlas3d/}. Their results show that $\sim$20 percent of all nearby ETGs outside the Virgo cluster are surrounded by a low-column-density \hi \ disc with typical size of many tens of kpc, much larger than the stellar body. Some of these \hi \ discs are thought to originate from the accretion of gas-rich satellites as indicated by the misalignment of their angular momentum vectors with respect to those of the stellar discs \citep{2014MNRAS.444.3388S}. Some of the kinematically aligned gas discs may originate from the cooling of hot gas in the halo \citep[][]{2015MNRAS.451.1212N}. Whatever their origin, this paper is concerned with the impact of these large \hi \ discs on the properties of the host ETGs: do these \hi \ discs host any star formation (SF), and if so what impact does this have on the morphology and stellar populations of the host galaxy?
\begin{table*}
\caption{General properties of the \hi-rich sample}
\begin{center}
\begin{tabular}{lccccccccccc}
\hline
\hline
No	&Name	& $D$	&$V_{hel}$	&$M_{K}$	&log$_{10}$($R_{eff}$)	&$P.A.$	&$\varepsilon$	&$E(B-V)$	&log$_{10}M$(\hi)	&log$_{10}M_{\star, r}$\\
	&	&[Mpc]	&[km/s]	&[mag]	&[arcsec]	&[degrees]	& [ ]	&[mag]	&[M$_{\odot}$]	&[M$_{\odot}$]\\
	& (1)	&(2)	&(3)	&(4)	&(5)		&(6)	&(7)&		(8)	&(9)	&(10)\\
\hline
1	&NGC~2594	&35.1	&2362	&-22.36	&0.82	&306	&0.122	&0.050	&8.91	&10.47\\
2	&NGC~2685	&16.7	&875	&-22.78	&1.41	&37	&0.402	&0.051	&9.33	&10.31\\
3	&NGC~2764	&39.6	&2706	&-23.19	&1.09	&173	&0.218	&0.034	&9.28	&10.64\\
4	&NGC~2859	&27.0	&1690	&-24.13	&1.43	&269	&0.272	&0.017	&8.46	&10.97\\
5	&NGC~3414	&24.5	&1470	&-23.98	&1.38	&135	&0.079	&0.021	&8.28	&11.11\\
6	&NGC~3522	&25.5	&1228	&-21.67	&1.01	&241	&0.504	&0.020	&8.47	&10.31\\
7	&NGC~3619	&26.8	&1560	&-23.57	&1.42	&72	&0.020	&0.013	&9.00	&10.91\\
8	&NGC~3941	&11.9	&930	&-23.06	&1.40	&189	&0.338	&0.018	&8.73	&10.34\\
9	&NGC~3945	&23.2	&1281	&-24.31	&1.45	&159	&0.435	&0.023	&8.85	&11.02\\
10	&NGC~4036	&24.6	&1385	&-24.40	&1.46	&272	&0.507	&0.019	&8.41	&11.16\\
11	&NGC~4203	&14.7	&1087	&-23.44	&1.47	&206	&0.153	&0.012	&9.15	&10.60\\
12	&NGC~4262	&15.4	&1375	&-22.60	&1.10	&216	&0.427	&0.031	&8.69	&10.48\\
13	&NGC~4278	&15.6	&620	&-23.80	&1.50	&50	&0.230	&0.023	&8.80	&11.08\\
14	&NGC~5103	&23.4	&1273	&-22.36	&1.02	&108	&0.278	&0.015	&8.57	&10.29\\
15	&NGC~5173	&38.4	&2424	&-22.88	&1.01	&261	&0.214	&0.024	&9.33	&10.42\\
16	&NGC~5631	&27.0	&1944	&-23.70	&1.32	&34	&0.198	&0.017	&8.89	&10.89\\
17	&UGC~03960	&33.2	&2255	&-21.89	&1.24	&108	&0.645	&0.040	&7.79	&10.39\\
18	&UGC~09519	&27.6	&1631	&-21.98	&0.87	&198	&0.402	&0.018	&9.27	&10.06\\
\hline
\hline
\end{tabular}
\label{table:HIrich}
\end{center}
\begin{tablenotes}[para,flushleft]\footnotesize
Note.$-$ Column (1): The name is the principal designation from LEDA \citep{2003A&A...412...45P}. Column (2): distance in Mpc \citet{2011MNRAS.413..813C}. Column (3): heliocentric velocity \citet{2011MNRAS.413..813C}. Column (4): total galaxy absolute magnitude derived from the apparent magnitude in K~band \citet{2011MNRAS.413..813C}. Column (5): projected half-light effective radius \citet{2011MNRAS.413..813C}. Column (6): position angle of the galaxy calculated from the \hi \ image \citet{2014MNRAS.444.3388S}. Column (7): ellipticity of the galaxy calculated from the \hi \ image \citet{2014MNRAS.444.3388S}. Column (8): estimates of Galactic dust extinction \citet{2011ApJ...737..103S}. Column (9): total \hi \ mass calculated assuming galaxy distances given in column~(2) \citet{2012MNRAS.422.1835S}. Column (10): stellar mass calculated by using a mass-to-light ratio and a total luminosity in the r-band \citep{2013MNRAS.432.1709C}.
\end{tablenotes}
\end{table*}

The relation between star formation (SF), \hi \ and colour for different types of galaxies has been known for decades \citep[see, e.g.,][and references in]{1994ARA&A..32..115R}. In late-type galaxies we know that the \hi \ content is an important driving factor of the star formation history. For example, \citet{2009MNRAS.396L..41H} show that \hi \ deficient spiral galaxies are at least 1 mag redder in UV-infrared colours than gas-rich ones, suggesting that quenching of SF and depletion of cold gas happen together. Furthermore, \citet{2011MNRAS.412.1081W} find that late-type galaxies with an \hi \ content larger than average have bluer outer regions. Similarly, \citet{2014ApJ...793...40H} report that most of the massive \hi \ galaxies $-$ mostly blue spirals$-$ detected in the ALFALFA survey have strong colour gradients, being bluer in the outer regions. These blue colours indicate existence of young stellar populations, and therefore, a link between the presence of \hi \ and relatively recent SF. Such link is known to exist in the \hi-dominated outer regions of spirals and in dwarf galaxies \citep[e.g.,][]{2008AJ....136.2846B, 2010AJ....140.1194B,2011AJ....142...37S,2012A&A...545A.142B,2015MNRAS.449.3700R}.

Similar to late-type galaxies, evidence of SF is found in the outer regions of ETGs \citep{2009AJ....137.5037D, 2010ApJ...714L.290S, 2010ApJ...714L.171T, 2011ApJ...733...74L, 2012ApJ...755..105S}. For example, \citet{2012ApJ...755..105S} find extended UV discs in 76 percent of their ETGs (mostly UV rings with diameters of tens of kpc). In addition, \citet{2012ApJ...745...34M} find that $\sim$~42 percent of their ETGs with stellar mass between $10^{8}~\mathrm{and}~4\times10^{10}$ M$_{\odot}$ are UV-bright. These objects contain \hi \ and have blue UV-infrared colours together with enhanced SF outside one effective radius. Their results also support the idea that galaxy evolution can proceed from early- to late-type as discussed by, e.g., \citet{2009MNRAS.400.1225C}, \citet{2012ApJ...761...23F} and \citet{2012ApJ...755..105S}.

In this paper, we aim to further explore the spatially-resolved link between \hi \ and SF in the outer regions of ETGs by analysing the UV, optical and \hi \ images of galaxies in the S12 sample as well as a control sample of \hi-poor control ETGs. For this purpose we use the \hi \ images published by S12, UV imaging from the Galaxy Evolution Explorer (GALEX) and optical imaging from the Sloan Digital Sky Survey (SDSS). In Section 2, we describe the \hi-rich and -poor sample as well as their selection criteria. In section 3, we describe the data (\hi, FUV, and optical imaging) and the methodology used for our work. In particular, we study the effect of the GALEX point spread function (PSF) on our results. In Section 4, we discuss the UV and optical properties of ETGs in our sample as a function of \hi \ content. In section 5, we compare star formation rate (SFR) with the studies in the literature. We also compare the integrated SF and SF efficiency in the \hi \ disc with that of late-type and dwarf galaxies. In Section 6, we present the conclusions of our work.

\section{Sample}
\label{sec:sample}
We have selected our samples based on S12. They analysed the \hi \ content of 166 ETGs largely based on data from the WSRT. \hi \ is detected in $\sim$1/3 of the ETGs, which allows us to define both an \hi-rich and an \hi-poor (i.e., undetected) control sample. The latter is important to investigate whether the SF properties at the outskirts of ETGs depend on the presence of \hi. Namely, we select 24 \hi -rich galaxies with extended discs and 55 \hi-poor control galaxies.

We have removed 5 \hi-rich galaxies (NGC~3626, NGC~3838, NGC~3998, NGC~5582, UGC~06176) for which there are no available UV data and one more \hi-rich galaxy with unusable UV and no SDSS data (NGC~6798). Table \ref{table:HIrich} lists all the remaining 18 \hi-rich galaxies and their basic properties. We note that $-$except for NGC~4262$-$ all the galaxies are outside the Virgo cluster. 

The \hi-poor control galaxies are selected from the non-detections in the survey of S12. For each \hi-rich galaxy we select up to 4 \hi-poor control galaxies with properties as close as possible to the ones of the \hi-rich object.~Namely:\\

\noindent \textbf{i)} stellar mass $M_{\star}$ \citep{2013MNRAS.432.1709C} within +/- 0.8 dex (but typically within +/- 0.5 dex);\\
\noindent \textbf{ii)} environment density $\Sigma_{3}$ \citep{2011MNRAS.416.1680C}\footnote{Mean surface density of galaxies inside a cylinder of height $h$ = 600~km~s$^{-1}$ (i.e. $\Delta V_{hel}<$~300 km~s$^{-1}$) centred on the galaxy which contains the 3 nearest neighbours.} within +/- 0.8 dex (but typically within +/- 0.5 dex);\\
\noindent \textbf{iii)} Virgo cluster (non-) membership \citep{2011MNRAS.413..813C};\\
\noindent \textbf{iv)} kinematical classification \citep[fast- or slow rotator;][]{2011MNRAS.414..888E};\\
\noindent \textbf{v)} distance \citep{2011MNRAS.413..813C} within +/- 15 Mpc;\\
\noindent \textbf{vi)} similar GALEX FUV imaging exposure time.\\

\begin{figure*}
\includegraphics[scale=1.0]{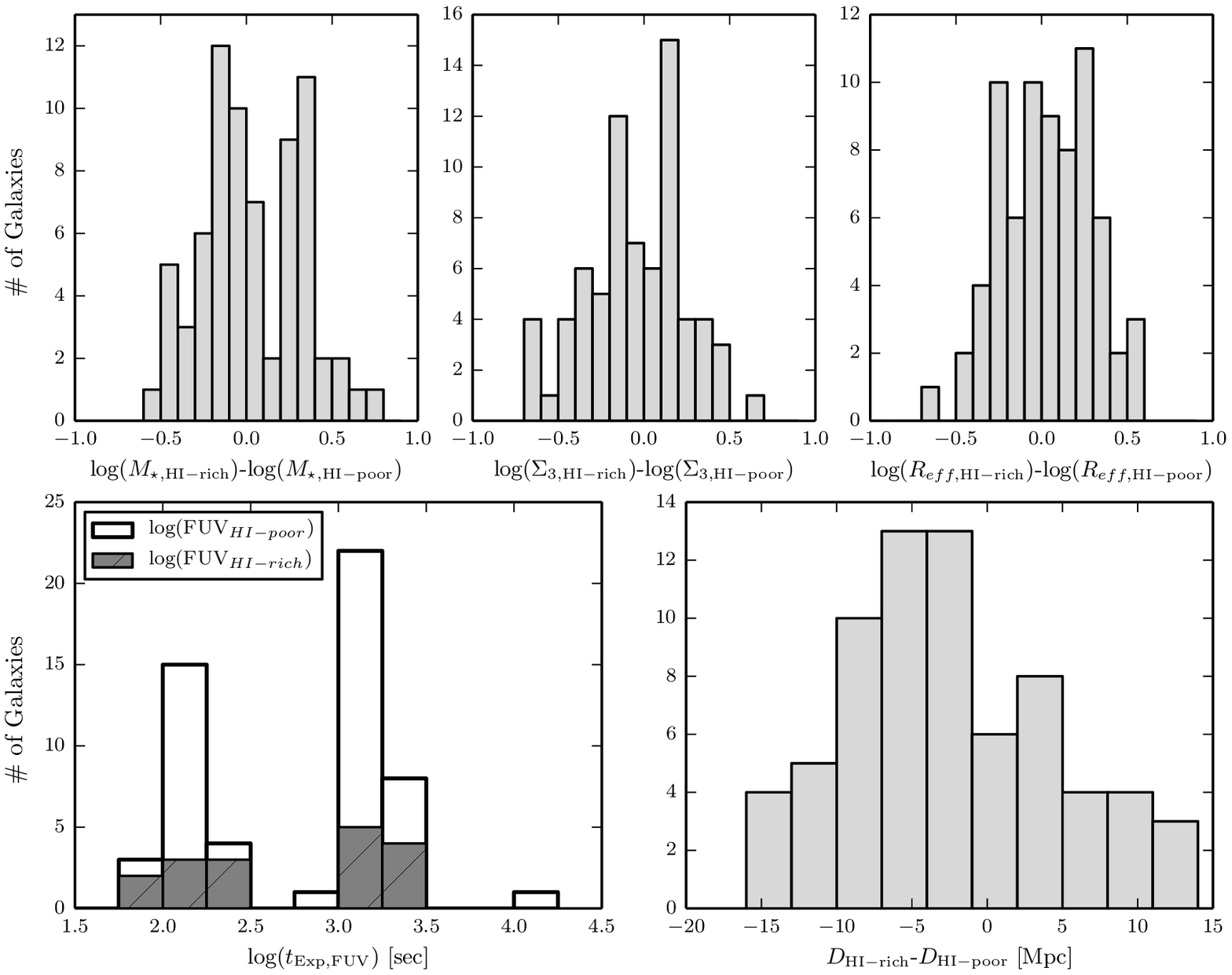}
\caption{\it Top panels. \rm Distribution of the difference between the stellar mass (left), environment density (middle) and effective radius (right) of the \hi-rich galaxies and \hi-poor control galaxies. \it Bottom panels. \rm FUV exposure time distribution of \hi-rich and -poor control galaxies (left) and distribution of the difference between the distance of the \hi rich galaxies and the one of their \hi-poor control galaxies. See section \ref{sec:sample} for details.}
\label{fig:histogram}
\end{figure*}

The resulting control galaxies are listed in Table \ref{table:HIcontrol}. The histograms in Fig. \ref{fig:histogram} compare \hi-rich and control samples. Apart for the quantities used for the selection of the control sample, which are by construction distributed similarly in the two samples, we note that also the distributions of effective radii $R_\mathrm{eff}$ are consistent with one another.
\section{Data reduction and analysis}
\label{sec:DATA_REDUC}
All galaxies (except 4 \hi-poor control galaxies) studied in this paper were observed with the WSRT. Fourteen \hi-rich galaxies were observed as a part of \atlas \ project; for the remaining 4 \hi-rich galaxies (NGC~2685, NGC~3414, NGC~4262, NGC~4278), we use \hi \ images published by \citet{2006MNRAS.371..157M}, \citet{2009A&A...494..489J}, \citet{2010MNRAS.409..500O}, while the $M$(\hi) values are taken from S12 in order to adopt consistent distances. The details about the observations and \hi \ data reduction are given in S12. The 4 \hi-poor control galaxies (NGC~4267, NGC~4377, NGC~4503, NGC~4762) mentioned above were observed \citet{2005AJ....130.2598G} as part of the ALFALFA survey (see S12 for details).
\subsection{GALEX data reduction}
\label{sec:galex}
\begin{table}
\caption{Control sample galaxies for each \hi-rich galaxy}
\resizebox{\columnwidth}{!}{%
\begin{tabular}{l@{\hspace{1em}}c@{\hspace{1em}}c@{\hspace{1em}}c@{\hspace{1em}}c@{}}
\hline
\hline
\hi-rich	&Control 1	& Control 2	& Control 3	&Control 4\\
\hline
NGC~2594	&UGC~04551	&NGC~5611	&NGC~3595	&NGC~6547\\
NGC~2685	&NGC~2549	&NGC~2852	&NGC~5273	&NGC~3098\\
NGC~2764	&NGC~4078	&NGC~2592	&PGC~044433	&NGC~2577\\
NGC~2859	&NGC~4762	&NGC~3230	&NGC~0821	&NGC~3458\\
NGC~3414	&NGC~5322	&-	&-	&-\\
NGC~3522	&NGC~3796	&-	&-	&-\\
NGC~3619	&NGC~5308	&NGC~3658	&NGC~5342	&NGC~2950\\
NGC~3941	&NGC~3605	&NGC~3301	&NGC~3248	&NGC~4283\\
NGC~3945	&NGC~3674	&NGC~4143	&NGC~3648	&NGC~3377\\
NGC~4036	&NGC~3613	&NGC~6548	&NGC~3665	&UGC~08876\\
NGC~4203	&NGC~3245	&NGC~3757	&-	&-\\
NGC~4262	&NGC~4503	&NGC~4340	&NGC~4267	&NGC~4377\\
NGC~4278	&NGC~5485	&NGC~3610	&NGC~4346	&-\\
NGC~5103	&NGC~2679	&NGC~0770	&PGC~051753	&-\\
NGC~5173	&NGC~5475	&NGC~5379	&NGC~3400	&NGC~5500\\
NGC~5631	&NGC~0661	&-	&-	&-\\
UGC~03960	&PGC~050395	&-	&-	&-\\
UGC~09519	&NGC~6149	&NGC~7457	&IC~3631	&-\\
\hline
\hline
\end{tabular}%
}
\label{table:HIcontrol}
\end{table}
We have obtained the GALEX UV images from the Mikulski Archive for Space Telescopes for all the \hi-rich and control sample galaxies. We used data from the All-sky Imaging Survey (AIS), Medium Imaging Survey (MIS), Guest Investigator Program (GII), Nearby Galaxy Survey (NGS) and Calibration Imaging (CAI), whose goals and specifications are described in \citet{2005IAUS..216..221M}. GALEX has two bands: the far-UV (FUV) and the near-UV (NUV), and the field of view (FOV) is $\sim$ 1.2 degrees in diameter. The effective wavelengths for the FUV and NUV bands are 1516\AA \ and 2373\AA, with 4.5 arcsec and 6 arcsec resolution (FWHM), respectively. All the technical details can be found in \citet{2005ApJ...619L...7M, 2007ApJS..173..682M}.

In general, the GALEX exposures are not centred on galaxies in our sample and, therefore, we analyse a $1500\times1500$ arcsec$^{2}$ region centered on our galaxies. Background images can be obtained as a product of the GALEX pipeline. However, the background levels are not sufficiently well determined in the regions far from the center of the FOV, and therefore we need to calculate the background around our galaxies again. To do so, we use the same method explained in \citet{2015MNRAS.451..103Y}. In this method, having masked the background and foreground objects based on SDSS photometric catalogue, images are clipped iteratively, and finally a mean moving filter is used to estimate the background. Before the clipping process, we mask a large area (300 arcsec diameter) around the centre of all \hi-rich and control sample galaxies to avoid over-estimating the background. We subtract these re-calculated backgrounds from the original, masked images. All galaxies show some residual background emission after the subtraction. Therefore, we subtract a flat background estimated from the UV radial profiles (Sec. \ref{sec:rad_pro}).

As a last step, we correct the UV images for the effect of Galactic extinction. We use $E$(B-V) values (see Table \ref{table:HIrich}) from \citet{2011ApJ...737..103S} based on the dust images by \citet{1998ApJ...500..525S} and assume $A_{\mathrm{FUV}}$=8.24~$\times~E$(B-V) and $A_{\mathrm{NUV}}$=8.2~$\times~E$(B-V) \citep{2007ApJS..173..293W}.

In this paper, the UV images are smoothed to the \hi \ resolution whenever a pixel-by-pixel comparison is made with the \hi \ data. For the other cases such as a colour calculation, we use the standard resolution of the UV images. During the reduction process, we use the natural unit of the UV images, which is counts per second (CPS). We convert this unit to magnitude as in \citet{2007ApJS..173..682M} by using the following equations :\\

FUV: m$_{AB}$ = $-$2.5 $\times$ log$_{10}$(CPS) + 18.82\\
\indent NUV: m$_{AB}$ = $-$2.5 $\times$ log$_{10}$(CPS) + 20.08\\

Having converted the CPS values to magnitudes, we apply the pixel size correction and convert the magnitudes to $mag~arcsec^{-2}$ by adding 0.8805 (the pixel size of the UV images is 1.5 arcsec).

Having subtracted the residual background from the UV images, we calculate the uncertainty in determining the background by using the same area used for the calculation of the residual background. We use at least 10 ellipses in the mentioned area and determine the background uncertainty as the standard deviation of the weighted mean intensities in the annuli between the ellipses. These uncertainties are used to calculate the error bars in the figures. 

\subsection{SDSS data}
\label{sec:sdss}
We make use of SDSS imaging data in order to gain a better understanding of the origin of the UV emission associated with the \hi. In our paper, we use archival SDSS -g and -r band mosaic images from DR10. We mask the SDSS images by using the same method explained above. Although the SDSS images have a flat background, generally a small residual background has to be subtracted. To determine the residual background, we use a single large elliptical annulus with a minimum inner radius of 100 arcsec and sufficiently far away (at least 20~R$_{eff}$) from the galaxy. We use the same ellipticity and position angle used for the radial profiles (see Sec. \ref{sec:rad_pro}). We calculate the residual background by taking the median pixel value within this annulus in the masked image. We also apply a correction for Galactic extinction by assuming $A_{\mathrm{SDSS-g}}$=3.793~$\times~E$(B-V) and $A_{\mathrm{SDSS-r}}$=2.751~$\times~E$(B-V) \citep{2002AJ....123..485S}.

When comparing the SDSS images with the UV, we convolve them with the FUV and NUV PSFs that are provided by the GALEX team. Since the convolution of the SDSS images with the UV PSFs creates some extended emission around the saturated objects, we have added some extra masks on these objects. For both the $g$ and $r$ band images, we convert the SDSS flux to magnitudes by using the following equation :\\\\
m$_{AB}$ = $-$2.5 $\times$ log$_{10}$(nanomaggies) + 22.50

As for the UV magnitudes, to obtain $mag~arcsec^{-2}$, we subtract 1.9897 from the SDSS magnitudes (the pixel size of the SDSS images is 0.4 arcsec). We also calculate the error bars with the same method used for the UV images.

\subsection{PSF effects}
\label{sec:psf_effect}

Since the GALEX PSF has considerable power at large radius \citep[][see the GALEX technical documentation\footnote{http://www.galex.caltech.edu/researcher/techdoc-ch5.html}]{2007ApJS..173..682M}, a strong central galaxy source, like a bulge, can contaminate the emission at larger radius. Therefore, the effect of the GALEX PSF is almost never negligible. In order to understand this effect we use a similar method as the one adopted by \citet{2012ApJ...745...34M}: after zeroing all the pixels outside 1 R$_{eff}$, the UV images are convolved with the GALEX PSF. Comparing the resulting images with the original ones, we can estimate how much of the light outside 1 R$_{eff}$ is due to the broad wings of the GALEX PSF.

In practice, having convolved the central UV images with the PSF, we quantify the PSF contamination by using radial profiles. Below we focus on two apertures, 1-3 and 3-10 R$_{eff}$, and here we verify whether any of these apertures is contaminated by the emission from within 1 R$_{eff}$ because of the large GALEX PSF. If the contamination from the central source is more than 20 percent in the first aperture, we increase the inner radius by the pixel size of the image. We repeat this process until the PSF contamination falls below 20 percent. In case the contamination is more than 20 percent everywhere in the first aperture, we flag this aperture as bad, and then continue the process for the second aperture (3 - 10 R$_{eff}$). With this method, we define a secure radius limit outside which the PSF contamination is less than 20 percent. In this paper, we use the secure radius limit for the next steps such as calculation of colour or star formation efficiency. We list the value of this radius for each galaxy in Table \ref{table:Colour_Flux}.
\begin{figure}
\includegraphics[scale=1.0]{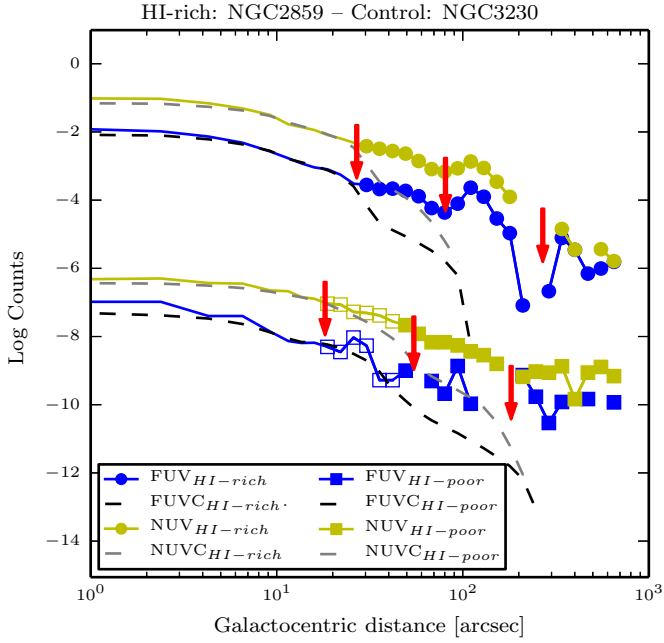}
\caption{Azimuthally averaged surface brightness profiles of the \hi-rich galaxy NGC~2859 (circles) and the \hi-poor control galaxy NGC~3230 (squares). For both galaxies we show the NUV and FUV profiles in yellow and blue, respectively. The profile of the PSF-convolved, central NUV and FUV emission is shown in grey and black, respectively. To separate the lines from two galaxies, the profiles of the control sample galaxy are scaled down by 5 dex. The red-vertical arrows correspond to 1, 3 and 10 effective radius, respectively. Open symbols indicate annuli below the secure radius, where the PSF contamination is large (Sec. \ref{sec:psf_effect}).}
\label{fig:psf_effect}
\end{figure}

Fig. \ref{fig:psf_effect} shows, as an example, one of the \hi-rich galaxies, NGC~2859 (upper part) and it’s control galaxy, NGC~3230 (lower part). In this figure, the black and grey dashed lines show the radial profiles of the convolved central FUV and NUV emission, respectively. The observed profiles are obtained from masked and carefully background-subtracted images. Ignoring the effect of the PSF, one might naively conclude that the FUV flux between 1 and 3 R$_{eff}$ is significant in both galaxies. However, in NGC~3230 this flux is almost entirely due to the PSF, and our method allows us to identify such situations. This shows the importance of taking into account the effect of the PSF, especially when dealing with the GALEX data.

Recent studies have found that the effect of the SDSS PSF can be important at large radius. For example, at 40 arcsec this effect is $\sim$ 10$^{-6}$ in relative flux \citep{2014MNRAS.443.1433D, 2014A&A...567A..97S}. At a similar radius the GALEX PSF is $\sim$ 100 times stronger than the SDSS one. Since we select galaxy regions where the GALEX PSF does not significantly contaminate the measured fluxes, the effect of the SDSS PSF should be negligible in the same regions.

\subsection{Measuring the SFR and \hi \ surface density}
\label{sec:measure}
The star formation rate (SFR) is estimated by using the FUV images described above. In order to convert the FUV emission into SFR, we follow \citet{2010AJ....140.1194B} and take:

\begin{equation}
\label{eq:eqSFR}
\Sigma_{\mathrm{SFR}}[\mathrm{M_{\odot}yr^{-1}kpc^{-2}}] = 0.68\times 10^{-28} \times I\mathrm{_{FUV}[erg\ s^{-1} Hz^{-1} kpc^{-2}]},
\end{equation}

\noindent where I$_{\mathrm{FUV}}$ is the FUV intensity per unit area and the initial mass function (IMF) is assumed to be of a Kroupa type. This estimate could in principle be corrected for internal extinction on the basis of the \hi \ image using the relation $N$(\hi)$/E(B-V)= 5 \times10^{21}$ cm$^{-2}$mag$^{-1}$ presented by \citet{1978ApJ...224..132B}. This relation is derived for the Milky Way and therefore is likely to provide an upper limit on the extinction in metal poor environments. Since the \hi \ column density is low (a few times $10^{19}$cm$^{-2}$) in the outer regions of ETGs, the extinction decreases to a negligible level $\sim$ 1-2 percent. In this paper, we do not apply this type of correction.

In order to compare the SFR with the \hi \ column density, we calculate the gas surface density from the \hi \ images. We convert the given \hi \ flux in to the gas surface density by using the standard conversion:

\begin{equation}
\label{eq:HI}
\Sigma_{HI}~[M_{\odot}~\mathrm{pc}^{-2}] = 8840 \times \frac{F_{\mathrm{HI}}~[\mathrm{Jy~/~beam} \times \mathrm{km/s}]} {\mathrm{bmaj} \times \mathrm{bmin~[arcsec]}^{2}}
\end{equation}

\noindent where the \textit{bmaj} and \textit{bmin} are the major and minor axis size of the beam, respectively. F$_{HI}$ is the flux detected across all velocity channels.

\section{Results}
\label{sec:properties}

\subsection{Radial profiles}
\label{sec:rad_pro}

In order to investigate the relation between \hi \ and star formation in the outer regions of ETGs, we derive the optical and UV radial profiles of the \hi-rich objects using elliptical annuli whose PA and ellipticity are those of the \hi \ disc. These values are derived as part of the kinematical analysis presented by \citet[][see table \ref{table:HIrich}]{2014MNRAS.444.3388S}. For the \hi-poor control galaxies we adopt the photometric PA and ellipticity published by \citet{2011MNRAS.414.2923K}. In both cases the width of the annuli increases logarithmically from the inside out.
In this paper, we do not study the SF inside 1~R$_{eff}$, because contamination by the UV flux from the old stellar populations makes the interpretation very difficult there.

\begin{figure}
\includegraphics[scale=1.0]{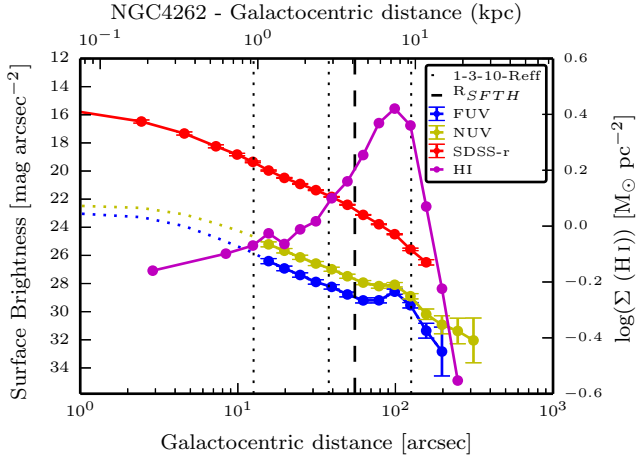}
\caption{Azimuthally averaged surface brightness profiles of the FUV (blue), NUV (yellow), SDSS-r band (red) images, and surface density profile of the \hi \ image (magenta) for NGC~4262. The \hi \ surface density units are given on the right-y axis in magenta. The black-vertical dashed line show the star formation threshold (see Sec. \ref{sec:rad_pro}). The vertical dotted lines corresponds to 1, 3 and 10 R$_{eff}$, respectively. The error bars represent the uncertainty in determining the background.}
\label{fig:rp_log}
\end{figure}

As an example, Figure \ref{fig:rp_log} shows the resulting \hi, UV and optical radial profiles for NGC~4262. In this figure, surface brightness values are converted to AB-mag arcsec$^{-2}$ and given on the left axis. The \hi \ surface density is given on the right axis. We provide a figure like Fig. \ref{fig:rp_log} for all \hi-rich galaxies in appendix \ref{App:AppendixA}.

As done in previous studies \citep[e.g.,][]{2007ApJS..173..538T,2011ApJ...733...74L,2012ApJ...745...34M}, we use the radius of the 27.35 mag~arcsec$^{2}$ NUV isophote as the star forming disc radius (R$_{\mathrm{SFTH}}$). Having calculated the R$_{\mathrm{SFTH}}$ values for all the \hi-rich galaxies, we find that all the \hi \ profiles extend well beyond this radius (e.g., see the black-vertical dashed line in Fig. \ref{fig:rp_log}). For NGC~4262, it is clear that the \hi \ disc not only extends much further out than the optical body but also that the highest \hi \ column density is found beyond this threshold. What is also interesting in Fig. \ref{fig:rp_log} is that the FUV and NUV profiles exhibit a local peak at the position of the peak of the \hi \ profile, whereas the r-band profile does not change. This region corresponds exactly to the star-forming UV ring almost 10~R$_{eff}$ away from the centre \citep[e.g.,][]{2009MNRAS.400.1225C}.

Based on the radial profiles, we find 9 \hi-rich galaxies for which the highest \hi \ column density region is beyond R$_{\mathrm{SFTH}}$. These are NGC~2594, NGC~2685, NGC~2859, NGC~3941, NGC~3945, NGC~4036, NGC~4262, NGC~4278, UGC~09519. The \hi \ column density of these 9 galaxies increases from the centre outwards, which means that the \hi \ is concentrated in the outer regions. Additionally and more importantly, in the majority of cases (but not all, such as, NGC~4278 \footnote{Although there are some UV clumps around NGC~4278, these are smoothed out in the radial profile due to the azimuthal averaging.}), the FUV and/or NUV profiles exhibit a slope change or a peak at the location of the \hi \ profile peak. This correspondence can be no coincidence and is a clear indication that the detected UV emission is related to SF triggered by the presence of \hi. This serves as a further demonstration that the outer regions of \hi-rich ETGs can be studied out to these very large radii using GALEX and SDSS data.
%
\begin{figure}
\includegraphics[scale=1.0]{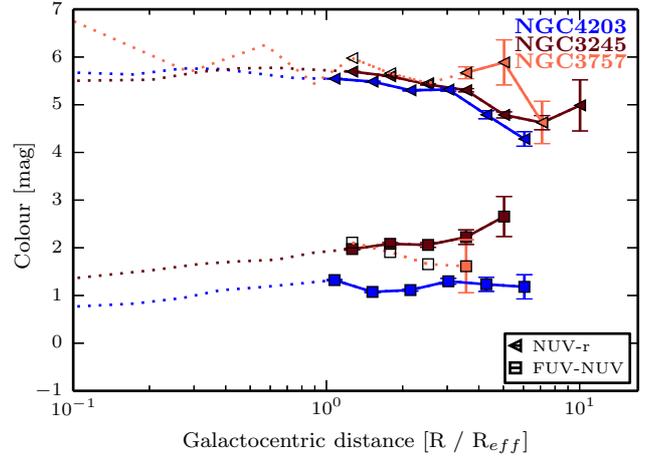}
\caption{\it Top: \rm Azimuthally averaged colour profiles for the \hi-rich galaxy NGC~4203 together with its control galaxies NGC~3245 and NGC~3757. The open symbols and the error bars are similar to those in Fig. \ref{fig:rp_log}. A different colour has been used for each galaxy, and this is indicated in the upper-right corner of the image.}
\label{fig:colour_colour}
\end{figure}

\subsection{UV-optical and UV-UV colour profiles}
\label{sec:col_pro}
Although the surface brightness profiles indicate where the UV and \hi \ emission coincides, they do not give a complete picture about the stellar populations. Besides, detecting an extended UV disc does not mean that the emission is coming from only the recently formed stars, the source of this emission could be mixture of young and old stellar populations. In order to understand this source, we also derive the $FUV-NUV$, $FUV-r$ and $NUV-r$ colour profiles for the \hi-rich and -poor control galaxies. We also show colour-colour diagrams for the \hi-rich and control sample galaxies. We note that our conclusions do not change if we use SDSS-g instead of SDSS-r band.

As an example, we show the colour profiles for NGC~4203 together with its control galaxies NGC~3245 and NGC~3757 in Fig. \ref{fig:colour_colour}. As can be seen from the colour profiles, NGC~4203 is generally bluer than its control galaxies, especially the colour $FUV-NUV$ \citep[see also][]{2015MNRAS.451..103Y}.

\subsection{Comparison of the two samples}
\label{sec:compare_samples}
\begin{table}
\begin{center}
\caption{The weighted average colour values together with their standard deviation.}
\scalebox{1.0}{
\begin{tabular}{lccccc}
\hline
\hline
Colour&$\overline{W}_{HI-rich}$&$\sigma$&$\overline{W}_{HI-poor}$&$\sigma$&$\Delta~\overline{W}_{R-P}$\\
\hline
FUV-NUV$_{Ap1}$&1.40&0.66&1.90&0.50&-0.50\\
FUV-NUV$_{Ap2}$&0.95&0.60&1.70&0.60&-0.75\\
FUV-r$_{Ap1}$&6.50&1.00&7.20&0.73&-0.70\\
FUV-r$_{Ap2}$&5.40&0.98&6.60&0.98&-1.20\\
FUV-g$_{Ap1}$&5.80&1.00&6.50&0.72&-0.70\\
FUV-g$_{Ap2}$&4.80&0.96&5.90&0.94&-1.10\\
NUV-r$_{Ap1}$&5.10&0.56&5.40&0.24&-0.30\\
NUV-r$_{Ap2}$&4.60&0.65&5.20&0.45&-0.60\\
NUV-g$_{Ap1}$&4.40&0.53&4.70&0.21&-0.30\\
NUV-g$_{Ap2}$&4.00&0.62&4.50&0.47&-0.50\\
\hline
\end{tabular}
}
\label{table:average_color}
\end{center}
\end{table}
As mentioned earlier, we have divided the outskirts of the galaxies into two region: 1-3 R$_{eff}$ and 3-10 R$_{eff}$. In this way, we can compare the colours of the \hi-rich and -poor control galaxies by integrating the light in the two regions. Additionally, we can compare the inner aperture with the outer one to determine the colour gradient in our galaxies.
\begin{figure*}
\includegraphics[scale=1.0]{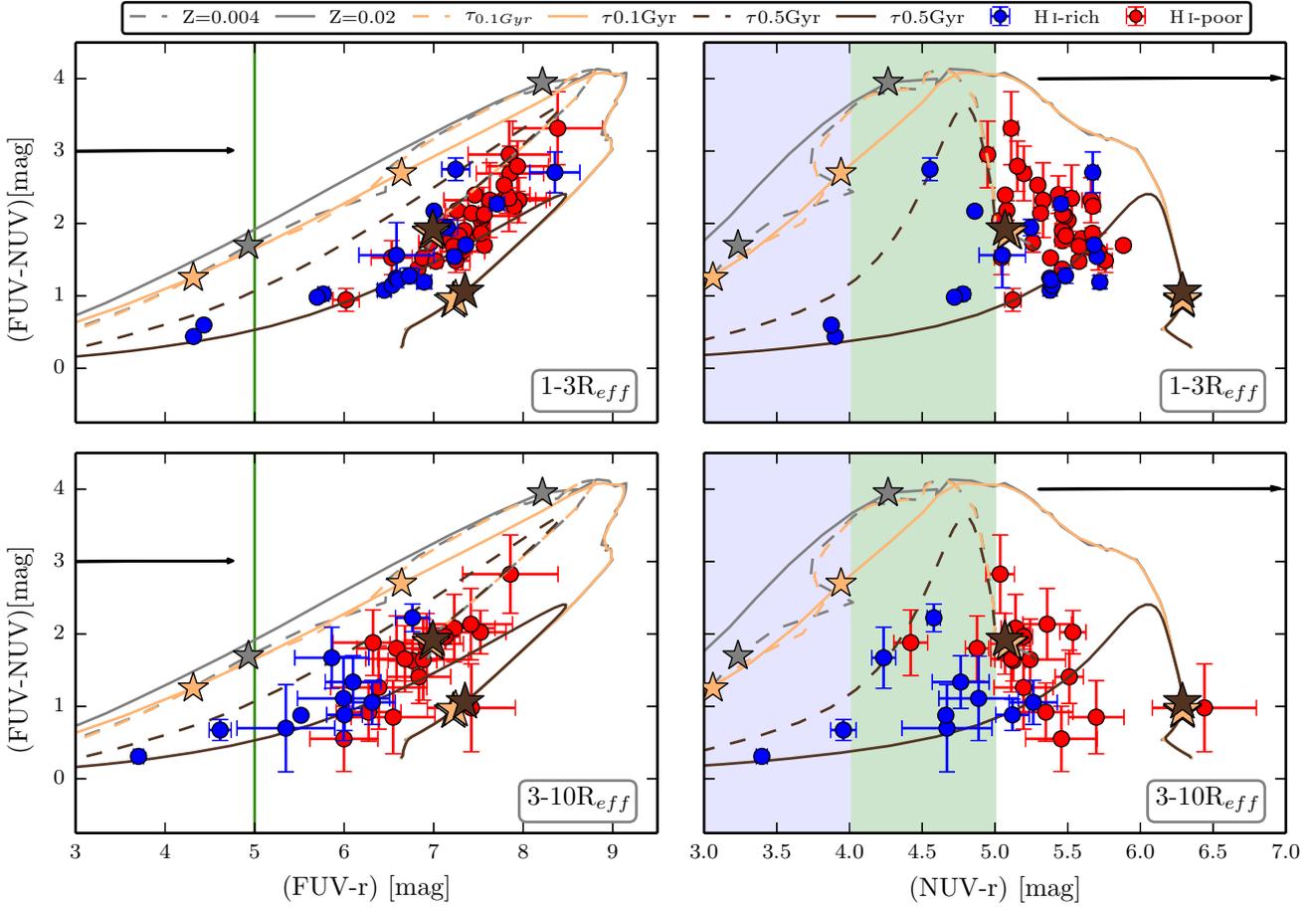}
\caption{Colour-colour diagrams for all the \hi-rich and -poor control galaxies in the sample. \it Top: \rm colour-colour diagrams for the aperture: 1~$-$~3~R$_{eff}$. \it Bottom panel: \rm colour-colour diagrams for the second aperture: 3~$-$~10~R$_{eff}$. The black arrow indicates the effect of an internal extinction of E(B-V)=0.3. Dashed and solid lines show the SSP models from \citet{2003MNRAS.344.1000B} for the metallicities of Z=0.004 and Z=0.02 (Solar), respectively. The grey and brown (light and dark) lines indicate instantaneous and exponentially declining (time scales are indicated in the legend) SF histories. The star shapes indicate the colour of the models at the age of 1 and 10 Gyr, and the sizes of the star shapes are in proportion to the ages. We note that the lines start to converge when the models become $\sim$~5~Gyr old. The models are almost completely converged at 10~Gyr, therefore, grey star shapes at this age stay behind the light-brown stars. The green line in the left panel shows the colour limit to indicate star-forming galaxies. The blue and green shaded areas in the right panel show the star-forming and transition galaxies, respectively (see Sec.~\ref{sec:stellar_pop}).}
\label{fig:link1}
\end{figure*}

We show all \hi-rich and control galaxies in Fig. \ref{fig:link1}. We list the weighted average of the colours, scatter and the colour differences between the \hi-rich and - poor control galaxies in table \ref{table:average_color}. This table and Fig. \ref{fig:link1} give the following results:\\

\noindent
(1) the overall average colour (all the colours used in this study) of the second aperture is 0.69 and 0.36 mag bluer than that of the first aperture for the \hi-rich and -poor control ETGs, respectively;\\
(2) the 6th column of the table shows that the \hi-rich ETGs are bluer in all colours than the \hi-poor control galaxies: 0.5 and 0.8 mag bluer in aperture 1 and 2, respectively.\\

To determine whether the distributions of the two samples are different from one other, we run a \textit{2D - 2 sample Kolmogorov-Smirnov} (K-S) test on our \hi-rich and -poor control galaxies. Table \ref{table:KS-test} lists the two colours used in the test and the resulting \textit{p}-value shows the probability that the two samples are from the same parent sample. The KS test is independent of the error bars of the measurements. In order to take these errors into account, we create 1000 synthetic points distributed as Gaussian around each point based on their error bars. We then run the \textit{2D - 2 sample K-S} test for the 1000 \hi-rich and -poor control pairs. Next to the p-values, we list the fraction of the trials that satisfies the 95 percent confidence level. As can be seen from the table, the q-values are smaller for the FUV-optical than the NUV-optical colours. It means that the difference between \hi-rich and -poor control ETGs becomes more clear for the FUV-optical colours. This result is consistent with our other result that the highest colour difference is found in the $FUV-r$ colour confirming the visual impression given by Fig. \ref{fig:link1}.
%
\begin{table}
\begin{center}
\caption{2D-KS test results}
\scalebox{1.0}{
\begin{tabular}{lcc}
\hline
\hline
Colour - Colour&\textit{p}$_{2D-KS test}$&$f_{trials}$~$<$~5~$\%$\\
\hline
FUV-r $-$ FUV-NUV$_{Ap1}$ &0.0020& 1.00\\
FUV-r $-$ FUV-NUV$_{Ap2}$ &0.0027& 0.98\\
NUV-r $-$ FUV-NUV$_{Ap1}$ &0.0069& 1.00\\
NUV-r $-$ FUV-NUV$_{Ap2}$ &0.0078& 0.82\\
\hline
\end{tabular}
}
\label{table:KS-test}
\end{center}
\end{table}

\begin{figure*}
\includegraphics[scale=1.00]{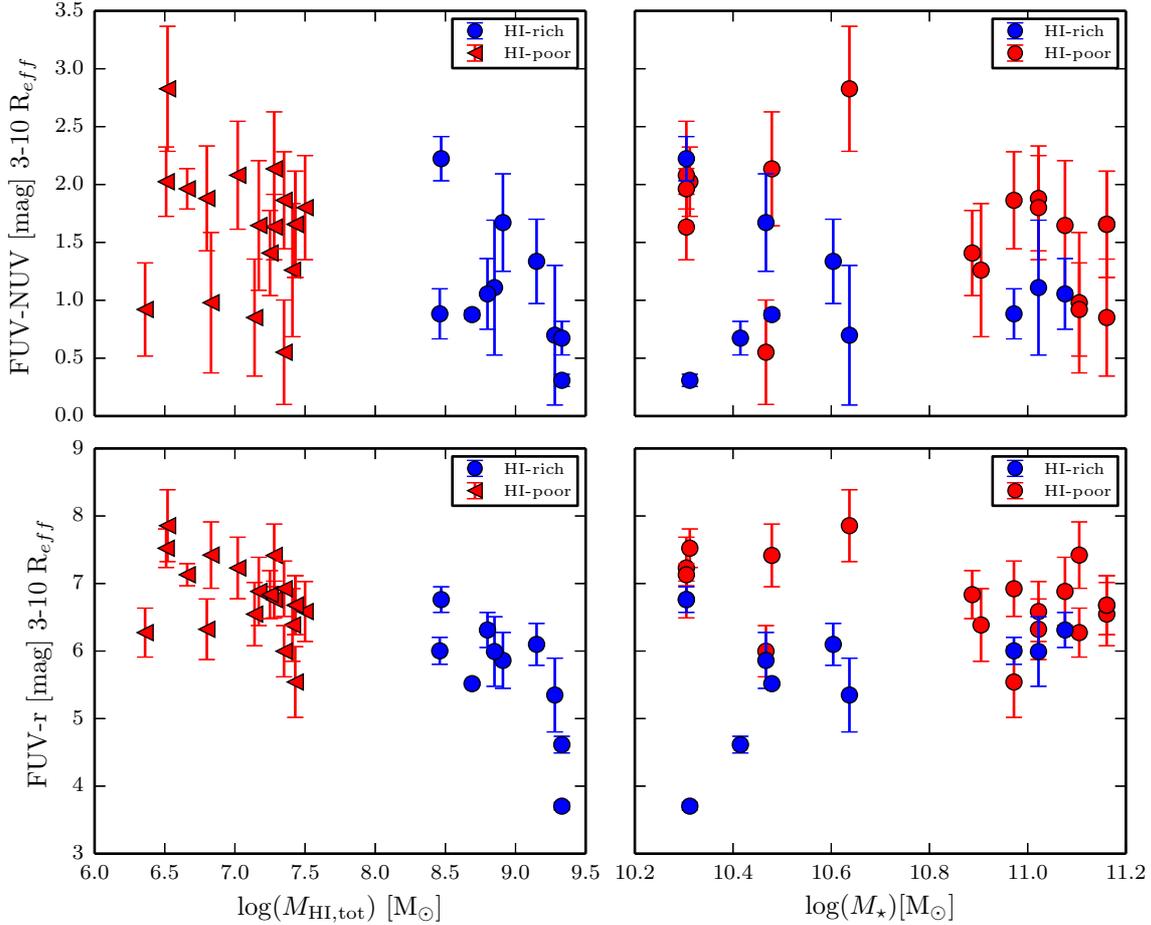}
\caption{Colour comparison of the \hi-rich and -poor control galaxies as a function of total \hi \ mass (\textit{left}) and stellar mass (\textit{right}). Note that \hi \ masses for the \hi-poor control galaxies are upper limits (see text).}
\label{fig:Colour_mass}
\end{figure*}

We also compare the colours of \hi-rich and -poor control galaxies at fixed stellar and \hi \ mass, and show the resulting comparison in Fig. \ref{fig:Colour_mass}. For the control galaxies we use the upper limits on $M$(\hi) published by S12. We find that in the $FUV-NUV$ colour \hi-rich and -poor control galaxies are slightly different from each other. A notable aspect of this figure is that the bluest galaxies in $FUV-r$ colour (bottom panel of Fig. \ref{fig:Colour_mass}) have the highest \hi-mass and the lowest stellar mass. The bottom-right panel of Fig. \ref{fig:Colour_mass} also shows that in a given stellar mass (below log$M_{\star}=$ 10.8) $FUV-r$ colour of the \hi-rich galaxies are bluer than the \hi-poor control ones supporting our findings mentioned earlier. However, there is almost no difference in colours for the stellar mass higher than log$M_{\star}=$ 10.8. 

\section{Discussion}
\subsection{Stellar populations}
\label{sec:stellar_pop}

Stellar populations of the \hi-rich and control sample galaxies are vital to understand the evolution of galaxies and, more importantly, the effect of the \hi \ on this evolution. As presented in Sec \ref{sec:compare_samples}, the \hi-rich ETGs have bluer outer regions than the \hi-poor control ETGs. Additionally, it is clear from table \ref{table:average_color} and Fig. \ref{fig:link1} that the scatter in colours is larger for the \hi-rich galaxies than for the \hi-poor control ones. This is likely due to the younger stellar populations in the outer regions of \hi-rich ETGs. \citet{2012ApJ...761...23F} similarly find that all the extended UV disc ETGs in their sample have bluer outer UV-optical colours.

The presence of young stellar populations in ETGs have been studied by several authors. For instance, \citet{2007ApJS..173..619K} find that galaxies with a $NUV-r$ colour less than 5.5 mag have most likely experienced recent SF. \citet{2014SerAJ.189....1S} also use a similar colour limit ($NUV-r=5.0$) as the criteria for transition region (i.e., green valley) galaxies. Moreover, \citet{2011ApJ...733...74L} use a colour limit at $FUV-r=5.0$ to separate evolved stellar populations from the recently formed stars in the extended-UV discs. We show these criteria and limits in Fig. \ref{fig:link1}.

It is striking that two \hi-rich ETGs, NGC~2685 and NGC~5173, show very blue outer regions (see Fig. \ref{fig:link1}). Within the aperture 1-3~R$_{eff}$ or 3-10~R$_{eff}$, the colours of these galaxies fall in the region occupied by star-forming late-type galaxies. In addition, based on their colour in the second aperture, 8 \hi-rich galaxies reside in a region bluer than $NUV-r=5$, indicating presence of young stellar populations. According to the colour of the second aperture, most of the \hi-rich galaxies reside below $FUV-r=6$ (see Fig. \ref{fig:Colour_mass}). \citep{2012ApJ...755..105S} have used this colour value (below $FUV-r=6$) as an indication of SF rather than old-star UV upturn phenomenon. In addition, these authors find that extended SF incidence rate starts to decrease after log$M_{\star}=10.9$ which might be related our findings that the colour difference between the \hi-rich and -poor control galaxies is lower for stellar masses above log$M_{\star}=10.9$.

Finally, our results are consistent with the stellar population models from \citet{2003MNRAS.344.1000B} \footnote{The SSP models used in our paper are calculated with the Kroupa type IMF and are updated in 2007.}: as can be seen in Fig. \ref{fig:link1}, the models with instantaneous SF history (grey lines) and exponentially declining SF history with a time scale of 0.1 Gyr (brown lines) show that 1 Gyr old stellar populations generally have $NUV-r$ colours between 3.0 and 4.5 regardless of metallicity. Even the exponentially declining SF history models with a time scale of 0.5 Gyr (dark brown lines) show relatively young stellar populations (up to 5~Gyr) between 3.0 and 4.5 $NUV-r$ colours.

\subsection{The link between \hi \ and SF}
We have established that the \hi-rich ETGs have bluer outskirts than the \hi-poor control ones, and these regions contain young stellar populations too. In this section, we use information from the surface brightness profiles to investigate the relation between the \hi \ content and SF in \hi-rich galaxies. We will also compare the blue outer discs of our ETGs with those of \hi \ dominated systems. To do so, we integrate the FUV flux, as well as the \hi \ flux in the two apertures used above and calculate the total SFR and total \hi \ mass.

\begin{figure}
\includegraphics[scale=1.0]{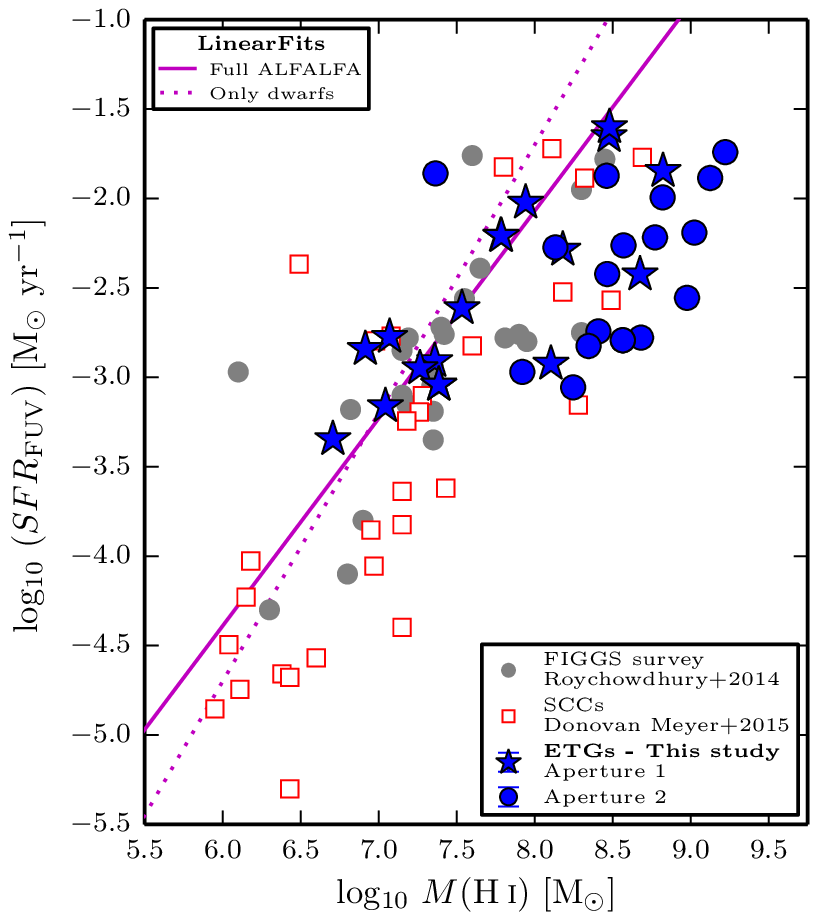}
\caption{Integrated SFR through the first and second aperture against the total \hi\ mass in the same apertures. SFR is calculated from the total FUV flux in the apertures. As comparison, we show dwarf irregular galaxies from \citet{2014MNRAS.445.1392R}, and single cloud candidates from \citet{2015ApJ...808..136D}. The SFR values of the comparison studies have obtained for the whole galaxies. We also show two linear fit lines: all the ALFALFA galaxies (solid magenta line) and only the dwarf galaxies (magenta dashed line) from \citet{2012ApJ...756..113H, 2012AJ....143..133H}.}
\label{fig:Integrated_SFR}
\end{figure}

Figure \ref{fig:Integrated_SFR} shows the resulting integrated SFR in the first (blue star shapes) and second (blue circles) apertures for the \hi-rich galaxies. In this figure, we also compare outer regions of \hi-rich ETGs with \hi-dominated systems such as faint dwarf irregular galaxies \citep[grey circles,][]{2014MNRAS.445.1392R} and dwarf candidate \hi \ clouds (red squares, these are detected in blind \hi \ observations and shown as \textit{dwarf candidates} based on their UV counterparts, see \citealt{2015ApJ...808..136D} for details).

The \hi \ mass in the second aperture is higher than in the first aperture for 16 \hi-rich ETGs (i.e., $\sim$ 89 percent). This includes objects where the \hi \ radial profile peaks within the second aperture, as well as objects where the \hi \ disc is very large and, therefore, most of its mass is outside 3 R$_{eff}$. As mentioned in Sec. \ref{sec:rad_pro}, there is a change in the UV profiles of 6 ETGs in our sample where the \hi-column density peaks (5 of these galaxies reside below $NUV-r = 5.0$). It is useful to note that these peaks are the azimuthally averaged values, thus they do not reflect the local gas surface density. For example, the UV profiles of NGC~4278 does not show a clear change where the \hi \ profile reaches its peak value $\sim0.7~\mathrm{M}_{\odot}~\mathrm{pc}^{-2}$ ($\sim$3 $R_{eff}$). However, at the same radius there are local \hi \ clumps with a column density of $\sim1.4~\mathrm{M}_{\odot}~\mathrm{pc}^{-2}$, and these are most likely associated with UV clumps indicative of localised star formation. Although the peak of \hi \ is not beyond the R$_{\mathrm{SFTH}}$, a similar case is also true for NGC~4203: \citet{2015MNRAS.451..103Y} show that some local high density \hi \ clumps can have SFR surface density as high as the inner regions of late-type galaxies, but this information may be lost when using radial profiles or quantities integrated over large apertures. 

We find the average \hi \ mass as $\sim1.4~\times10^{8}~\mathrm{M}_{\odot}$ and $\sim5\times10^{8}~\mathrm{M}_{\odot}$ for the first and second apertures, respectively. Although the \hi \ mass in the second aperture is 4 times higher than in the first aperture, the integrated SFR values are almost the same: $6.1\times10^{-3}~\mathrm{M}_{\odot}~\mathrm{yr}^{-1}$ for the first and $6.2~\times10^{-3}~\mathrm{M}_{\odot}~\mathrm{yr}^{-1}$  for the second aperture. This indicates a lower SF efficiency in the outer aperture, as we discuss in the next Section.
\subsection{SF efficiency and comparison with the literature}
\label{sec:comparing_SFR}
In the previous sections, based on the good match between features in \hi \ and UV radial profiles, we conclude that the \hi-rich ETGs have blue outskirts as a consequence of low-level SF, triggered by the presence of \hi. In this section, we discuss SFR surface density as a function of \hi \ column density. We also investigate how efficiently the \hi \ is converted into new stars in these regions, and how this compares to the situation in other galaxies.
\begin{figure}
\includegraphics[scale=1.0]{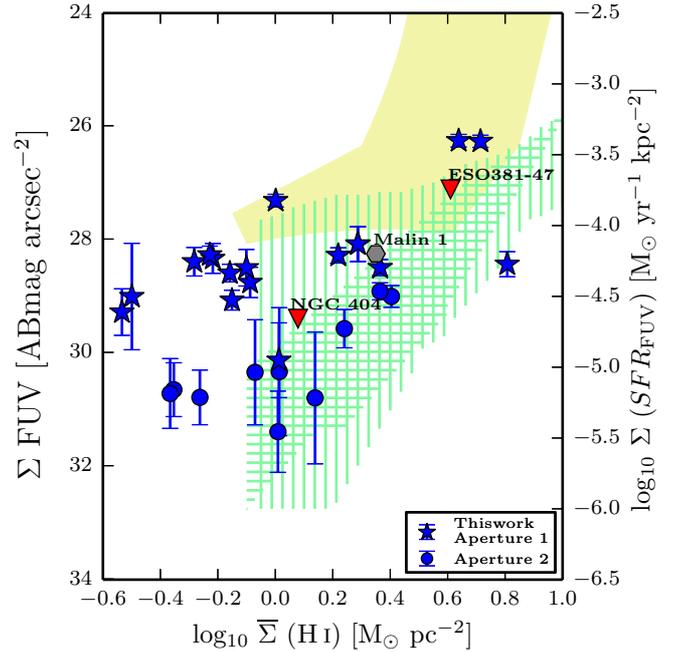}
\caption{SFR - \hi \ relation as a function of radius for the \hi-rich galaxies. The star shapes and circles indicate the average SFR and \hi \ column density of the aperture 1 and aperture 2, respectively. The yellow area shows the relation between \hi \ and SFR in the sample of late-type galaxies of  \citet{2008AJ....136.2846B}. The vertical and horizontal green-shaded areas show the location of dwarf galaxies and outer regions of spirals, respectively \citep{2010AJ....140.1194B}. In this Figure, the left axis shows the FUV surface brightness and the right axis shows the SFR surface density calculated with Eq. \ref{eq:eqSFR}.}
\label{fig:ellbyell}
\end{figure}

By using Eq. \ref{eq:eqSFR} and \ref{eq:HI}, we calculate the SFR- and \hi- surface density ($\Sigma_{SFR}$ and $\Sigma_{HI}$) in the first and second aperture. The resulting plot is shown in Fig. \ref{fig:ellbyell}. An important characteristic of this figure is that the average $\Sigma_{SFR}$ in the first aperture is almost 1 dex higher than in the second aperture: $\Sigma_{\mathrm{SFR,Ap1}}\approx 9.1 \times 10^{-5}~\mathrm{M}_{\odot}~\mathrm{yr}^{-1}~\mathrm{kpc}^{-2}$ and $\Sigma_{\mathrm{SFR,Ap2}}\approx 1.3 \times 10^{-5}~\mathrm{M}_{\odot}~\mathrm{yr}^{-1}~\mathrm{kpc}^{-2}$ for the first and second apertures, respectively. Table \ref{table:SFR-HIvalue} is showing the SFR surface density values in the aperture 1 and 2 for all the \hi-rich sample galaxies.

We compare our results with those of \citet{2008AJ....136.2846B, 2010AJ....140.1194B} who studied the same relationship for the inner regions of late-type galaxies and the outer regions of spirals and dwarf galaxies (shown in Fig. \ref{fig:ellbyell} as yellow shaded area, green vertical and horizontal lines, respectively). It is striking that the SFR surface density levels in the outer regions of ETGs are comparable with those of outer regions of spiral and dwarf galaxies. Moreover, the SFR surface density levels of NGC~2685 and NGC~5173 in the first aperture are the same as those of the inner regions of late type galaxies. Since our first aperture is between 1-3 R~$_{eff}$, even this region can be counted as an outer region. It is worth to note that these galaxies are the bluest galaxies in our sample and their colour is similar to those of star-forming galaxies. In Fig. \ref{fig:ellbyell}, we show additional data points (red triangles) for two lenticular galaxies: NGC~404 and ESO~381-47. The values are calculated for the UV/\hi \ rings far from the optical body of the host galaxies by \citet{2010ApJ...714L.171T} and \citet{2009AJ....137.5037D}, respectively.

We also show a grey hexagon data point in Fig. \ref{fig:ellbyell} corresponding to a giant low-surface brightness galaxy (LSB): Malin~1. \citet{2010A&A...516A..11L} suggest that Malin~1 contains a double structure: an inner high surface brightness early-type spiral galaxy and an outer extended LSB disc. These authors find that the total \hi \ mass is $\sim~6.7\times10^{10}~\mathrm{M}_{\odot}$ and the disc extends up to $\sim$110~kpc radii. In addition, a recent deep optical study has revealed that Malin~1 extends up to 150 kpc \citep{2015ApJ...815L..29G}. Although the SFR values of Malin~1 are obtained for a large area including the central regions \citep[see][]{2009ApJ...696.1834W}, it is remarkable that our results are consistent with that of Malin 1. Since our \hi-rich galaxies have bright inner regions surrounded with inefficient star forming low-column density \hi \ discs, we can ask this question: are the \hi-rich ETGs related to giant LSB objects such as Malin-1? However, we need more information and analysis to answer this question, therefore we will leave this topic to future studies.

\begin{figure}
\includegraphics[scale=1.0]{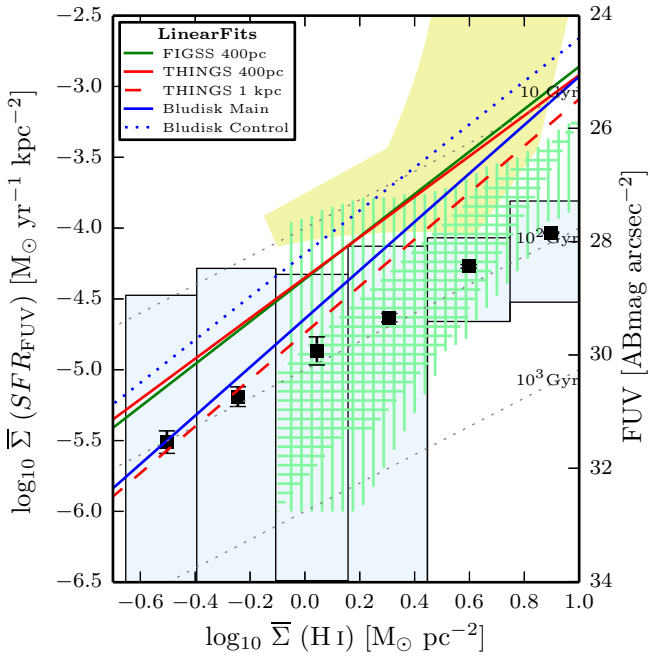}
\caption{$\Sigma$~SFR$_{FUV}$ as a function of \hi \ column density for all the \hi-rich galaxies in our sample (black squares with error bars). The error bars are the uncertainties in the background calculation. The light blue bars show the scatter of FUV pixels in the bins. We are using both the positive and negative FUV pixels in the bins, therefore, two blue bars go to the $-$infinity for the lowest two \hi \ bin. Here, we consider only the regions beyond 3$R_{eff}$. Grey dotted lines indicate various constant gas depletion time scales. The yellow and green shaded areas are the same as in Fig. \ref{fig:ellbyell}. Different colour lines show the same relation for different samples (see text). These lines are drawn by using the slope and intercept values taken from table 4 of \citet{2015MNRAS.449.3700R}.}
\label{fig:intermsofhi}
\end{figure}

So far we have compared \hi \ and SFR surface density within fixed apertures. Here we perform a pixel by pixel as done by previous authors \citep[e.g.,][]{2008AJ....136.2846B,2010AJ....140.1194B,2012A&A...545A.142B,2015MNRAS.451..103Y}. To do so, we smooth and re-grid the FUV images to the resolution and coordinate grid of the \hi \ images. We show the smoothed FUV images together with \hi-column density contours and 3 ellipses indicating 1-3-10~R$_{reff}$ in Fig. \ref{fig:appendix3}. We focus on regions outside 3~R$_{reff}$, and for galaxy we calculate the average FUV flux in bins of constant \hi \ column density. The bins have a width of 0.3 dex and are identical for all galaxies. For each bin we then take the weighted mean of all the single-galaxy average FUV fluxes. These weighted mean values are indicated by the black markers in Fig. \ref{fig:intermsofhi}, while the cyan boxes show the scatter of all pixels in each bin.

\begin{table*}
\begin{center}
\caption{Logarithmic \hi \ and star formation values of the \hi-rich galaxies in the aperture 1 and 2.}
\scalebox{1.0}{
\begin{tabular}{lcccccccccc}
\hline
\hline
Galaxy&\textbf{log$_{10}$~[}~$M$(\hi)$_{Tot,Ap1}$&$M$(\hi)$_{Tot,Ap2}$&$\Sigma_{HI,Ap1}$&$\Sigma_{HI,Ap2}$&$SFR_{FUV,Ap1}$&$SFR_{FUV,Ap2}$&$\Sigma_{SFR,Ap1}$&$\Sigma_{SFR,Ap2}$&$SFE_{Ap1}$&$SFE_{Ap2}$~\textbf{]}\\
&\multicolumn{2}{c}{[M$_{\odot}$]}&\multicolumn{2}{c}{[M$_{\odot}$~pc$^{-2}$]}&\multicolumn{2}{c}{[M$_{\odot}$~yr$^{-1}$]}&\multicolumn{2}{c}{[M$_{\odot}$~yr$^{-1}$~kpc$^{-2}$]}&\multicolumn{2}{c}{[yr$^{-1}$]}\\
\hline
NGC~2594& 6.71& 8.41& -0.53& -0.07& -3.3& -2.7& -4.6& -5.0& -10.1& -11.2\\
NGC~2685& 8.48& 9.22& 0.71& 0.40& -1.7& -1.7& -3.4& -4.5& -10.1& -11.0\\
NGC~2764& 8.68& 9.02& 0.81& 0.14& -2.4& -2.2& -4.3& -5.2& -11.1& -11.2\\
NGC~2859& --& 8.46& --& -0.26& -2.3& -1.9& -4.6& -5.2& --& -10.3\\
NGC~3414& 7.94& 7.36& -0.28& -0.72& -2.0& -1.9& --& -5.1& -10.0& --\\
NGC~3522& 7.04& 7.92& -0.16& -0.35& -3.2& -3.0& -4.3& -5.2& -10.2& -10.9\\
NGC~3619& 8.82& 8.18& 0.36& -0.13& -1.8& -2.3& -4.3& --& -10.7& --\\
NGC~3941& 7.36& 8.68& -0.09& 0.14& -2.9& -2.8& -4.4& --& -10.3& --\\
NGC~3945& 7.78& 8.82& -0.22& 0.01& -2.2& -2.0& -4.2& -5.0& -10.0& -10.8\\
NGC~4036& 7.53& 8.35& -0.15& -0.03& -2.6& -2.8& -4.5& --& -10.1& --\\
NGC~4203& 8.18& 8.98& 0.22& 0.01& -2.3& -2.6& -4.2& -5.5& -10.5& -11.5\\
NGC~4262& 7.07& 8.47& 0.00& 0.36& -2.8& -2.4& -3.8& -4.5& -9.8& -10.9\\
NGC~4278& 7.79& 8.57& -0.23& -0.37& -2.2& -2.3& -4.2& -5.2& -10.0& -10.8\\
NGC~5103& 7.26& 8.25& -0.10& -0.17& -2.9& -3.1& -4.3& --& -10.2& -11.3\\
NGC~5173& 8.48& 9.12& 0.64& 0.24& -1.6& -1.9& -3.4& -4.7& -10.1& -11.0\\
NGC~5631& 8.10& 8.78& 0.01& -0.15& -2.9& -2.2& -5.0& --& -11.0& --\\
UGC~03960& 6.91& 7.51& -0.50& -0.56& -2.8& --& -4.5& --& -9.8& --\\
UGC~09519& 7.39& 8.56& 0.29& 0.38& -3.0& -2.8& -4.1& --& -10.4& --\\
\hline
\end{tabular}
}
\label{table:SFR-HIvalue}
\end{center}
\begin{tablenotes}[para,flushleft]\footnotesize
Note.$-$ Column (1): The name of the \hi-rich galaxies. The values in the next columns are for the first and second apertures (1-3 and 3-10 R$_{eff}$), respectively. Column (2-3): total \hi \ mass. Column (4-5): average \hi \ column density. Column (6-7): integrated star formation rate. Column (8-9): average star formation density. Column (10-11): star formation efficiency.
\end{tablenotes}
\end{table*}

It is clear from Fig. \ref{fig:intermsofhi} that the $\Sigma_{SFR}$ level is increasing with increasing \hi\ column density (as seen also in Fig. \ref{fig:Integrated_SFR} and \ref{fig:ellbyell}). However, the SF efficiency remains low at all \hi \ column densities in the outermost regions (R$>$3~R$_{eff}$) shown in this figure: the average SF efficiency is $\sim~1.3~(0.3 - 5) \times10^{-11}$~yr$^{-1}$. This is $\sim$ 5 times lower than in aperture 1. The time necessary to consume all the \hi \ gas (without gas recycling) is $\sim~9~(1.5 - 23) \times~t_{H}$ (Hubble time) and $\sim~2~(0.7 - 9) \times~t_{H}$ for aperture 2 and 1, respectively.

As can be seen from Fig. \ref{fig:intermsofhi}, the outermost regions of early- and late-type galaxies are consistent with each others. The gas depletion time of the outer regions late-type galaxies is $\sim10^{11}$~yr ($\sim7\times~t_{H}$, \citet{2010AJ....140.1194B}). \citet{2015MNRAS.449.3700R} analyse the same relation between $\Sigma_{gas}$ and $\Sigma_{SFR}$ in the \hi \ dominated regions of different type of galaxies with a similar method to ours, and therefore we use their results (shown as colour lines in Fig. \ref{fig:intermsofhi}) to compare with ours. There is an offset between the lines of \citet{2015MNRAS.449.3700R} and our points, and this is most likely due to the fact that we explicitly select the outer regions of galaxies, while they include inner regions too (as long as they are \hi \ dominated). It is worth to note that, for our sample, the scatter in the $\Sigma_{SFR}$ is increasing with decreasing $\Sigma_{gas}$.

\subsection{Implication for the evolution of these galaxies}
\label{sec:link1}

From the UV-optical colours, we show that several galaxies are blue in the outer regions due to recent star formation related to the presence of \hi \ gas. Their outer colours are so blue that if they were total colours, we would consider those ETGs to lie in the green valley or even blue cloud. Are these outer regions affecting the evolution of the host galaxies? Another way of asking this question is: can the \hi \ gas in outer regions fuel the star formation at a sufficiently high level to rejuvenate them and change the morphology of the host by building a stellar disc? In this section, we will try to answer these questions.

As mentioned in the previous sections, NGC~2685 and NGC~5173 are the bluest galaxies in our sample (in both apertures)\footnote{If we integrate the flux of NGC~2685 and NGC~5173 up to 10 R$_{eff}$, they reside in the green valley and blue-cloud regions, respectively.}. The gas depletion time of these galaxies is $\sim$ 10$^{11}$~yr in the second aperture, accordingly, the outer regions of these galaxies will be bluer than the inner regions for a long time. However, given the low-level (inefficient) star formation, the \hi \ gas cannot change the morphology of the host galaxies. If we integrate the total SF over 1 Gyr (assuming constant SF), the contribution to the total stellar mass will be $\sim$ 0.1 and 0.05 percent for NGC~2685 and NGC~5173, respectively. Similarly, three \hi-rich galaxies, NGC~2594, NGC~2764, NGC~3522, show bluer colours than $NUV-r=5.0$ for both the aperture 1 and 2. These galaxies also can stay in these regions for a long time without changing the morphology of the host due to the inefficient SF.

There are three \hi-rich galaxies, NGC~3945, NGC~4203, NGC~4262, whose colour in the second aperture is much bluer than in the first (they are in the red-sequence even if we integrate their flux up to 10~Re). While the gas consumption time in the inner regions is relatively short, it increases dramatically in the second aperture (e.g., $\sim$6 Gyr and 79 Gyr for NGC~4262). On the one hand, if we integrate the total SF over 1 Gyr, it is not enough to create a disc-dominated system. On the other hand, outer regions of these galaxies can stay in the green valley region for a long time, while the inner regions stay in the red-sequence. However, \citet{2009MNRAS.400.1225C} note that if the flux is integrated up to the UV ring ($\sim$~13~R$_{eff}$), NGC~4262 moves to the transition region (i.e., green valley) in $NUV-K$ colour.

Although the average colour difference is very small (see Sec. \ref{sec:compare_samples}), there are some \hi-poor control galaxies have bluer outer regions. For example, the $NUV-r$ colour of NGC~4143 is almost 1 mag bluer in the second aperture than in the first and, therefore, the outermost region of this galaxy appears in the green valley region. Galaxies similar to NGC~4143 could be in the process of moving towards the red-sequence due to lack of gas as it is claimed in the general galaxy evolution idea. However, contrary to a general idea, we now know that the \hi-rich ETGs might have come from the red-sequence to the green valley or even if they are moving towards the redder regions they will move slowly. Our conclusions are supported by several studies in the literature \citep[e.g.,][]{2009MNRAS.400.1225C, 2012MNRAS.419.1727C}. For example, \citet{2009MNRAS.400.1225C}, show that while all the \hi-deficient spiral galaxies $-$mainly found in the Virgo cluster$-$ are moving towards the red-sequence, some galaxies lying between the red and blue sequence might have different evolutionary paths: moving into $-$ and staying temporarily in $-$ the transition region. In addition, \citet{2012ApJ...745...34M} show that their UV-bright ETGs have larger \hi \ reservoirs than the rest of their sample, and these galaxies are growing a stellar disc and are mostly located in the blue sequence. It is believed that extended and large \hi \ structures in ETGs are associated with external accretion \citep[e.g.,][]{2006MNRAS.371..157M, 2007A&A...465..787O, 2010MNRAS.409..500O}. Thus, these galaxies might have come from the red-sequence to the blue-cloud. A recent simulation by \citet{2016MNRAS.460.3925T} also supports this idea that some galaxies --have already moved to the red-sequence-- return to the blue sequence again by having recent star formation. They find that the majority of the blue galaxies were red in the past. They also claim that most of the galaxies stay in the green valley region at least $\sim$ 2 Gyr.  

As a result, if the gas in the outer regions of ETGs is converted to stars for 1 Gyr, the average new-to-total(already existing) stellar mass ratio becomes 1.6$\times$10$^{-4}$. This result indicates that at present the recent/new star formation does not have a significant impact on the structure of the host galaxies. However, the present \hi \ gas is enough to keep the outer regions of the \hi-rich galaxies in the green valley or blue-cloud for a long time. It is important to note that if these \hi-rich systems lose their gas supply due to exhaustion or some external effect such as merging, the evolution of these systems currently in the blue cloud or in transition will change: they will move towards the redder colours.

\section{Conclusions}
\label{sec:conclusions}

In this paper we have investigated the relation between \hi, star formation and colours of the outer regions of ETGs by comparing an \hi-rich sample to an \hi-poor one. We have used spatially resolved \hi \ images together with the GALEX UV and SDSS $g,r$ band images. We have used two apertures to study the outer regions: 1-3 and 3-10 R$_{eff}$. In addition to the fixed apertures, we also study the SF as a function of \hi \ column density. Our main conclusions are given below.

1. Beyond 1~R$_{eff}$, \hi-rich ETGs are bluer (in UV-optical colours) than \hi-poor control ETGs. This holds also at fixed stellar mass for $M_{\star}<6 \times 10^{10}~\mathrm{M}_{\odot}$. In some extreme ETGs the outer colour is comparable to that of late-type galaxies.\\
\indent2. The \hi-rich galaxies have much stronger colour gradients --between the first and second apertures-- than the \hi-poor control galaxies. This means that the presence of the \hi \ has led to star formation in the outer parts of ETGs.\\
\indent3. In $\sim$ 89 percent of the \hi-rich ETGs, the \hi \ mass is higher in the second aperture than in the first. On average the \hi \ mass in the second aperture is 5 times higher than in the first aperture.\\
\indent4. Ten \hi-rich galaxies show increasing \hi \ profiles. In relation to that the peak \hi \ column density for 9 of these \hi-rich ETGs is beyond the SF threshold radius. More importantly, in five of the cases, at the same position, the FUV or NUV profile shows a change in their slope.\\
\indent5. The SFR surface density in the in the first aperture is almost 1 dex higher than in the second aperture: 9.1 and $1.3 \times 10^{-5}~\mathrm{M}_{\odot}~\mathrm{yr}^{-1}~\mathrm{kpc}^{-2}$, respectively.\\
\indent6. We have found that outermost regions of 8 \hi-rich ETGs reside between the blue-cloud and red-sequence. This situation is likely to persist for a long time due to their low efficient SF: the average SF efficiency of the second aperture is $1\times10^{-11}$~yr$^{-1}$.

Although ETGs are quite different from late-type or dwarf galaxies in their central regions, they are very similar when considering the \hi \ dominated outer regions. Early-types, spirals and dwarfs all show a similar $\Sigma$~SFR vs $\Sigma$~\hi \ relation, which shows that SFR is increasing with increasing \hi \ surface density. Additionally, they show similarly low SF efficiency. The gas depletion time for the outermost regions of ETGs and late-type galaxies is almost the same: $\sim10^{11}$~yr. Another meaning of these results could be that forming stars in a low column density region does not depend on the host galaxy type.

\section{Acknowledgements}

MKY acknowledges a Ph.D. scholarship from The Council of Higher Education of Turkey and is supported by the University of Erciyes. We wish to thank  Alessandro Boselli, Alexandar Bouquin, Jennifer Donovan Meyer, Luca Cortese, and Manolis Papastergis for useful discussions and helpful comments on the manuscript.

We would like to thank the GALEX team for their great work to make the data public and available. This work uses observations made with the NASA Galaxy Evolution Explorer. GALEX is operated for NASA by Caltech under NASA contract NAS5-98034. Part of this work is based on observations made with the Spitzer Space Telescope, which is operated by the Jet Propulsion Laboratory(JPL), California Institute of Technology under a contract with NASA. This research has made use of the NASA/IPAC Extragalactic Database (NED) which is operated by JPL, Caltech, under contract with NASA.
The optical images used in this paper were obtained from the SDSS; funding for the SDSS and SDSS-II has been provided by the Alfred P. Sloan Foundation, the Participating Institutions, the National Science Foundation, the U.S. Department of Energy, the National Aeronautics and Space Administration, the Japanese Monbukagakusho, the Max Planck Society, and the Higher Education Funding Council for England. The SDSS Web site is http://www.sdss.org/.

\bibliography{biblio.firstmky.bib}

\appendix
\section{\\Appendix A}
\label{App:AppendixA}
\pagestyle{empty}
\begin{landscape}
\begin{table}
\caption{The limits of the aperture 1 (R1-R2) and 2 (R2-R3), total flux in different bands, and the colour values in the first and second apertures for the \hi-rich and -poor control galaxies.}
\scalebox{0.9}{
\begin{tabular}{lc|ccc|cccc|cccc|ccccc|cccccc}
\hline
\hline
&&\multicolumn{3}{c|}{Secure Radius}&\multicolumn{4}{c|}{Integrated Flux in Aperture1}&\multicolumn{4}{c|}{Integrated Flux in Aperture2}&\multicolumn{5}{c|}{Average colour in Aperture1}&\multicolumn{5}{c}{Average colour in Aperture2}\\
Galaxy$_{HIrich}$&$R_{eff}$&$R1$&$R2$&$R3$&$FUV$&$NUV$&$r$&$g$&$FUV$&$NUV$&$r$&$g$&$FUV-NUV$&$FUV-r$&$FUV-g$&$NUV-r$&$NUV-g$&$FUV-NUV$&$FUV-r$&$FUV-g$&$NUV-r$&$NUV-g$\\
&[arcsec]&\multicolumn{3}{c|}{[R$_{eff}$]}&\multicolumn{4}{c|}{[mag]}&\multicolumn{4}{c|}{[mag]}&\multicolumn{5}{c|}{[mag]}&\multicolumn{5}{c}{[mag]}\\
\hline
\hline
NGC~2594	& 6.61	& 1.7	& 3.0	& 10.0	& 22.28	& 19.53	& 18.06	& 18.75	& 20.77	& 19.14	& 18.03	& 18.75	& 2.75	& 7.25	& 6.57	& 4.55	& 3.87	& 1.67	& 5.86	& 5.18	& 4.23	& 3.54\\
NGC~2685	& 25.70	& 1.0	& 3.0	& 10.0	& 16.43	& 15.99	& 15.05	& 15.71	& 16.66	& 16.36	& 15.86	& 16.53	& 0.44	& 4.32	& 3.65	& 3.90	& 3.23	& 0.31	& 3.70	& 3.10	& 3.40	& 2.80\\
NGC~2764	& 12.30	& 1.9	& 3.0	& 10.0	& 20.24	& 19.21	& 17.40	& 17.98	& 19.66	& 18.97	& 17.15	& 17.59	& 1.03	& 5.77	& 5.18	& 4.78	& 4.19	& 0.70	& 5.35	& 4.90	& 4.67	& 4.22\\
NGC~2859	& 26.92	& 1.0	& 3.0	& 10.0	& 19.13	& 17.59	& 14.82	& 15.53	& 18.03	& 17.16	& 14.73	& 15.35	& 1.54	& 7.23	& 6.51	& 5.70	& 4.98	& 0.88	& 6.00	& 5.38	& 5.12	& 4.50\\
NGC~3414	& 23.99	& 1.0	& 3.0	& 10.0	& 18.19	& 17.14	& 14.64	& 15.38	& 17.79	& 18.08	& 15.03	& 15.76	& 1.08	& 6.45	& 5.71	& 5.38	& 4.64	& --	& --	& --	& --	& --\\
NGC~3522	& 10.23	& 1.3	& 3.0	& 10.0	& 21.12	& 18.98	& 17.05	& 17.71	& 20.65	& 18.45	& 16.87	& 17.49	& 2.17	& 7.00	& 6.34	& 4.86	& 4.19	& 2.22	& 6.76	& 6.11	& 4.58	& 3.93\\
NGC~3619	& 26.30	& 1.0	& 3.0	& 10.0	& 17.94	& 16.96	& 15.15	& 15.82	& 19.01	& 17.60	& 15.66	& 16.31	& 0.98	& 5.70	& 5.02	& 4.72	& 4.04	& --	& --	& --	& 4.73	& 4.02\\
NGC~3941	& 25.12	& 1.0	& 3.0	& 10.0	& 18.84	& 16.57	& 14.06	& 14.76	& 18.51	& 17.48	& 15.38	& 15.97	& 2.27	& 7.71	& 7.00	& 5.45	& 4.74	& --	& --	& --	& 4.97	& 4.34\\
NGC~3945	& 28.18	& 1.2	& 3.0	& 10.0	& 18.54	& 17.29	& 14.52	& 15.29	& 18.00	& 16.79	& 14.73	& 15.44	& 1.19	& 6.89	& 6.13	& 5.72	& 4.95	& 1.11	& 5.99	& 5.29	& 4.89	& 4.19\\
NGC~4036	& 28.84	& 1.7	& 3.0	& 10.0	& 19.68	& 17.98	& 15.27	& 15.99	& 20.21	& 18.38	& 15.45	& 16.09	& 1.70	& 7.36	& 6.62	& 5.68	& 4.95	& --	& --	& --	& 5.79	& 5.07\\
NGC~4203	& 29.51	& 1.0	& 3.0	& 10.0	& 17.74	& 16.60	& 14.11	& 14.84	& 18.42	& 17.10	& 15.19	& 15.56	& 1.14	& 6.53	& 5.80	& 5.39	& 4.66	& 1.34	& 6.10	& 5.70	& 4.76	& 4.36\\
NGC~4262	& 12.59	& 1.0	& 3.0	& 10.0	& 19.06	& 17.81	& 15.47	& 16.23	& 18.18	& 17.34	& 15.68	& 16.36	& 1.25	& 6.58	& 5.81	& 5.38	& 4.61	& 0.88	& 5.52	& 4.82	& 4.66	& 3.96\\
NGC~4278	& 31.62	& 1.0	& 3.0	& 10.0	& 17.68	& 16.46	& 14.00	& 14.73	& 17.81	& 16.79	& 14.34	& 15.05	& 1.22	& 6.60	& 5.87	& 5.39	& 4.66	& 1.06	& 6.31	& 5.63	& 5.26	& 4.58\\
NGC~5103	& 10.47	& 1.1	& 3.0	& 10.0	& 20.41	& 18.47	& 16.18	& 16.87	& 20.68	& 19.20	& 16.55	& 17.16	& 1.94	& 7.15	& 6.46	& 5.25	& 4.55	& --	& --	& --	& 5.36	& 4.78\\
NGC~5173	& 10.23	& 1.0	& 3.0	& 10.0	& 18.11	& 17.53	& 16.64	& 17.25	& 18.83	& 18.18	& 17.23	& 17.79	& 0.60	& 4.43	& 3.81	& 3.88	& 3.25	& 0.67	& 4.61	& 4.05	& 3.96	& 3.39\\
NGC~5631	& 20.89	& 1.2	& 3.0	& 10.0	& 20.66	& 17.96	& 15.24	& 15.95	& 18.89	& 17.79	& 15.35	& 16.04	& 2.71	& 8.35	& 7.63	& 5.67	& 4.95	& --	& --	& --	& 5.26	& 4.62\\
UGC~03960	& 17.38	& 1.0	& 3.0	& 10.0	& 20.90	& 19.34	& 17.21	& 17.89	& --	& 18.95	& 16.76	& 17.41	& 1.56	& 6.58	& 5.89	& 5.05	& 4.36	& --	& --	& --	& --	& --\\
UGC~09519	& 7.41	& 1.2	& 3.0	& 10.0	& 21.00	& 19.72	& 17.23	& 17.96	& 20.38	& 19.58	& 17.75	& 18.50	& 1.28	& 6.73	& 5.99	& 5.48	& 4.74	& --	& --	& --	& 4.93	& 4.19\\
\hline
\hline
&&\multicolumn{3}{c|}{Secure Radius}&\multicolumn{4}{c|}{Integrated Flux in Aperture1}&\multicolumn{4}{c|}{Integrated Flux in Aperture2}&\multicolumn{5}{c|}{Average colour in Aperture1}&\multicolumn{5}{c}{Average colour in Aperture2}\\
Galaxy$_{HIpoor}$&$R_{eff}$&$R1$&$R2$&$R3$&$FUV$&$NUV$&$r$&$g$&$FUV$&$NUV$&$r$&$g$&$FUV-NUV$&$FUV-r$&$FUV-g$&$NUV-r$&$NUV-g$&$FUV-NUV$&$FUV-r$&$FUV-g$&$NUV-r$&$NUV-g$\\
&[arcsec]&\multicolumn{3}{c|}{[R$_{eff}$]}&\multicolumn{4}{c|}{[mag]}&\multicolumn{4}{c|}{[mag]}&\multicolumn{5}{c|}{[mag]}&\multicolumn{5}{c}{[mag]}\\
\hline
NGC~0661	& 13.18	& 1.3	& 3.0	& 10.0	& 21.17	& 18.85	& 16.15	& 16.90	& 22.13	& 18.75	& 16.07	& 16.78	& 2.32	& 7.95	& 7.20	& 5.66	& 4.91	& --	& --	& --	& --	& --\\
NGC~2549	& 19.05	& 2.8	& 3.0	& 10.0	& 23.31	& 21.07	& 18.31	& 19.04	& 19.66	& 17.66	& 15.13	& 15.84	& 2.24	& 7.90	& 7.17	& 5.67	& 4.94	& 2.02	& 7.52	& 6.81	& 5.54	& 4.82\\
NGC~2679	& 22.39	& 1.0	& 3.0	& 10.0	& 21.08	& 18.69	& 16.20	& 16.89	& 20.50	& 20.44	& 17.70	& 18.38	& 2.40	& 7.82	& 7.12	& 5.44	& 4.74	& --	& --	& --	& --	& --\\
NGC~3245	& 25.12	& 1.2	& 3.0	& 10.0	& 19.35	& 17.31	& 14.74	& 15.44	& 20.38	& 17.44	& 15.42	& 16.02	& 2.05	& 7.53	& 6.82	& 5.50	& 4.79	& --	& --	& --	& 4.91	& 4.28\\
NGC~3605	& 16.98	& 1.3	& 3.0	& 10.0	& 20.86	& 19.09	& 16.62	& 17.30	& 20.97	& 19.99	& 16.33	& 17.03	& 1.76	& 7.18	& 6.51	& 5.46	& 4.79	& 0.98	& 7.42	& 6.75	& 6.44	& 5.77\\
NGC~3613	& 26.30	& 1.2	& 3.0	& 10.0	& 19.38	& 17.59	& 14.95	& 15.68	& 18.97	& 17.35	& 15.03	& 15.71	& 1.79	& 7.36	& 6.63	& 5.60	& 4.86	& 1.63	& 6.76	& 6.07	& 5.13	& 4.44\\
NGC~3674	& 11.22	& --	& 3.0	& 10.0	& --	& --	& 16.03	& 16.78	& 20.34	& 19.09	& 17.05	& 17.76	& --	& --	& --	& --	& --	& 1.26	& 6.39	& 5.66	& 5.20	& 4.46\\
NGC~3796	& 11.48	& 1.7	& 3.0	& 10.0	& 22.17	& 19.48	& 17.27	& 17.92	& 21.32	& 19.84	& 17.29	& 17.93	& 2.69	& 7.85	& 7.20	& 5.20	& 4.55	& --	& --	& --	& --	& --\\
NGC~4078	& 8.32	& 2.4	& 3.0	& 10.0	& --	& 20.79	& 18.58	& 19.27	& 21.27	& 18.98	& 16.84	& 17.31	& --	& --	& --	& 5.33	& 4.66	& --	& --	& --	& 5.16	& 4.64\\
NGC~4503	& 28.18	& 1.3	& 3.0	& 10.0	& 19.58	& 17.89	& 14.93	& 15.70	& 19.13	& 18.31	& 15.58	& 16.33	& 1.70	& 7.57	& 6.79	& 5.88	& 5.11	& 0.85	& 6.55	& 5.80	& 5.70	& 4.95\\
NGC~5322	& 39.81	& 1.0	& 3.0	& 10.0	& 18.59	& 16.73	& 13.97	& 14.70	& 18.20	& 16.92	& 14.07	& 14.82	& 1.86	& 7.53	& 6.80	& 5.67	& 4.94	& --	& --	& --	& 5.67	& 4.94\\
NGC~5475	& 16.60	& --	& 4.5	& 10.0	& --	& --	& 15.38	& 16.12	& 21.76	& 19.66	& 17.52	& 18.17	& --	& --	& --	& --	& --	& 2.14	& 7.42	& 6.74	& 5.36	& 4.68\\
NGC~5485	& 28.18	& 1.0	& 3.0	& 10.0	& 18.91	& 17.55	& 15.00	& 15.74	& 19.22	& 17.59	& 15.09	& 15.83	& 1.37	& 6.82	& 6.08	& 5.46	& 4.72	& 1.65	& 6.88	& 6.14	& 5.24	& 4.50\\
NGC~6149	& 10.72	& 1.3	& 3.0	& 10.0	& 21.79	& 19.46	& 17.07	& 17.78	& 20.22	& 19.75	& 17.52	& 18.24	& 2.32	& 7.63	& 6.92	& 5.33	& 4.63	& --	& --	& --	& --	& --\\
PGC~05039	& 10.96	& 1.4	& 3.0	& 10.0	& 23.46	& 20.15	& 18.03	& 18.70	& 22.03	& 20.19	& 18.43	& 19.06	& 3.32	& 8.39	& 7.71	& 5.11	& 4.44	& --	& --	& --	& --	& --\\
UGC~04551	& 10.72	& --	& 3.0	& 10.0	& --	& --	& 16.05	& 16.81	& 20.65	& 19.26	& 16.93	& 17.69	& --	& --	& --	& --	& --	& 1.41	& 6.84	& 6.05	& 5.51	& 4.73\\
NGC~0770	& 8.71	& 1.7	& 3.0	& 10.0	& 23.05	& 20.23	& 17.98	& 18.69	& 20.40	& 19.17	& 17.30	& 18.02	& --	& --	& --	& 5.32	& 4.60	& --	& --	& --	& 4.76	& 4.05\\
NGC~2592	& 12.30	& 1.0	& 3.0	& 10.0	& --	& 18.65	& 16.06	& 16.82	& --	& 19.34	& 17.15	& 17.94	& --	& --	& --	& 5.56	& 4.80	& --	& --	& --	& 5.29	& 4.52\\
NGC~2852	& 7.08	& 1.8	& 3.0	& 10.0	& 21.38	& 19.85	& 17.85	& 18.50	& 20.09	& 19.09	& 17.38	& 17.88	& 1.53	& 6.53	& 5.86	& 5.05	& 4.37	& --	& --	& --	& 4.74	& 4.18\\
NGC~3230	& 18.20	& 2.5	& 3.0	& 10.0	& 22.57	& 20.22	& 17.68	& 18.47	& --	& 18.42	& 16.43	& 17.16	& 2.35	& 7.84	& 7.07	& 5.53	& 4.76	& --	& --	& --	& 5.14	& 4.44\\
NGC~3301	& 19.95	& 2.8	& 3.0	& 10.0	& 22.13	& --	& 18.02	& 18.70	& 19.81	& 17.66	& 15.51	& 16.19	& --	& 7.03	& 6.35	& --	& --	& --	& --	& --	& 5.16	& 4.47\\
NGC~3610	& 15.85	& 1.0	& 3.0	& 10.0	& 19.60	& 17.07	& 14.76	& 15.43	& 19.64	& 17.57	& 15.28	& 15.97	& 2.53	& 7.79	& 7.12	& 5.29	& 4.62	& 2.08	& 7.23	& 6.54	& 5.14	& 4.45\\
NGC~3658	& 19.05	& 1.0	& 3.0	& 10.0	& 20.20	& 18.30	& 15.78	& 16.50	& 19.46	& 19.76	& 17.32	& 17.79	& 1.90	& 7.34	& 6.61	& 5.46	& 4.73	& --	& --	& --	& --	& --\\
NGC~3757	& 8.91	& --	& 3.0	& 10.0	& --	& --	& 15.44	& 16.17	& 20.36	& 19.32	& 16.75	& 17.45	& --	& --	& --	& --	& --	& --	& --	& --	& --	& --\\
NGC~4143	& 24.55	& 1.2	& 3.0	& 10.0	& 19.02	& 17.49	& 15.10	& 15.83	& 19.69	& 17.83	& 16.12	& 16.69	& 1.52	& 6.88	& 6.14	& 5.38	& 4.64	& 1.88	& 6.32	& 5.70	& 4.42	& 3.80\\
NGC~4340	& 37.15	& 1.0	& 3.0	& 10.0	& 19.17	& 17.56	& 14.74	& 15.49	& 21.94	& 18.76	& 15.80	& 16.69	& 1.61	& 7.32	& 6.57	& 5.71	& 4.96	& --	& --	& --	& --	& --\\
NGC~5611	& 10.00	& 2.4	& 3.0	& 10.0	& 23.12	& 20.17	& 18.30	& 19.01	& 20.02	& 19.29	& 17.23	& 17.92	& 2.95	& 7.84	& 7.14	& 4.95	& 4.25	& --	& --	& --	& 5.21	& 4.51\\
NGC~7457	& 36.31	& 1.1	& 3.0	& 10.0	& 19.69	& 16.91	& 14.65	& 15.33	& --	& 18.05	& 15.66	& 16.26	& 2.79	& 7.94	& 7.26	& 5.15	& 4.47	& --	& --	& --	& 5.36	& 4.76\\
\hline
\end{tabular}
\label{table:Colour_Flux}
}
\end{table}
\end{landscape}
\pagestyle{plain}

\pagestyle{empty}
\begin{landscape}
\begin{table}
\caption{The limits of the aperture 1 (R1-R2) and 2 (R2-R3), total flux in different bands, and the colour values in the first and second apertures for the \hi-poor control galaxies.}
\scalebox{0.9}{
\begin{tabular}{lc|ccc|cccc|cccc|ccccc|cccccc}
\hline
\hline
&&\multicolumn{3}{c|}{Secure Radius}&\multicolumn{4}{c|}{Integrated Flux in Aperture1}&\multicolumn{4}{c|}{Integrated Flux in Aperture2}&\multicolumn{5}{c|}{Average colour in Aperture1}&\multicolumn{5}{c}{Average colour in Aperture2}\\
Galaxy$_{HIpoor}$&$R_{eff}$&$R1$&$R2$&$R3$&$FUV$&$NUV$&$r$&$g$&$FUV$&$NUV$&$r$&$g$&$FUV-NUV$&$FUV-r$&$FUV-g$&$NUV-r$&$NUV-g$&$FUV-NUV$&$FUV-r$&$FUV-g$&$NUV-r$&$NUV-g$\\
&[arcsec]&\multicolumn{3}{c|}{[R$_{eff}$]}&\multicolumn{4}{c|}{[mag]}&\multicolumn{4}{c|}{[mag]}&\multicolumn{5}{c|}{[mag]}&\multicolumn{5}{c}{[mag]}\\
\hline
IC~3631	& 13.49	& 2.7	& 3.0	& 10.0	& 21.29	& --	& 20.23	& 20.86	& 19.79	& --	& 18.90	& 19.35	& --	& 4.09	& 3.50	& --	& --	& --	& --	& --	& --	& --\\
NGC~3248	& 15.85	& 1.0	& 3.0	& 10.0	& 20.16	& 18.12	& 16.03	& 16.68	& 20.28	& 18.05	& 16.28	& 16.91	& 2.04	& 7.05	& 6.40	& 5.03	& 4.38	& --	& --	& --	& 4.65	& 4.04\\
NGC~3400	& 16.98	& 1.2	& 3.0	& 10.0	& 21.52	& 19.29	& 16.81	& 17.52	& 19.95	& 20.20	& 18.00	& 18.74	& --	& --	& --	& 5.43	& 4.72	& --	& --	& --	& --	& --\\
NGC~3595	& 14.13	& 1.1	& 3.0	& 10.0	& 20.82	& --	& 16.11	& 16.80	& 19.36	& --	& 16.63	& 17.00	& --	& 7.61	& 6.91	& --	& --	& --	& 5.54	& 5.05	& --	& --\\
NGC~3648	& 13.18	& 2.4	& 3.0	& 10.0	& 22.52	& 20.42	& 17.94	& 18.65	& 20.39	& --	& 17.46	& 18.18	& 2.10	& 7.54	& 6.82	& 5.48	& 4.75	& --	& --	& --	& --	& --\\
NGC~3665	& 30.90	& 1.0	& 3.0	& 10.0	& 18.81	& 17.13	& 14.26	& 15.02	& 18.87	& 17.72	& 14.45	& 15.18	& 1.68	& 7.32	& 6.56	& 5.65	& 4.89	& --	& --	& --	& 5.88	& 5.16\\
NGC~4267	& 38.02	& 1.0	& 3.0	& 10.0	& 18.92	& 17.45	& 14.57	& 15.32	& 18.10	& 18.75	& 16.10	& 16.66	& 1.49	& 7.25	& 6.49	& 5.76	& 5.01	& --	& --	& --	& --	& --\\
NGC~5273	& 37.15	& 1.0	& 3.0	& 10.0	& 19.53	& 17.36	& 15.17	& 15.86	& 19.95	& 18.63	& 16.72	& 17.73	& 2.18	& 7.26	& 6.57	& 5.08	& 4.39	& --	& --	& --	& --	& --\\
NGC~5342	& 9.33	& 2.1	& 3.0	& 10.0	& --	& 20.42	& 18.17	& 18.89	& 21.28	& 19.57	& 17.81	& 18.53	& --	& --	& --	& 5.25	& 4.52	& 1.80	& 6.59	& 5.86	& 4.88	& 4.15\\
PGC~04443	& 5.13	& 1.0	& 3.0	& 10.0	& 21.21	& --	& 17.01	& 17.79	& 21.23	& 19.72	& 17.62	& 18.31	& --	& 6.77	& 6.00	& --	& --	& --	& --	& --	& 5.34	& 4.62\\
NGC~2577	& 14.13	& 1.1	& 3.0	& 10.0	& 20.02	& 18.54	& 15.92	& 16.66	& 19.37	& 18.84	& 16.28	& 16.99	& 1.48	& 7.02	& 6.28	& 5.58	& 4.83	& 0.55	& 6.00	& 5.26	& 5.46	& 4.72\\
NGC~2950	& 15.49	& 1.2	& 3.0	& 10.0	& 19.37	& 17.24	& 14.88	& 15.55	& 19.28	& 17.45	& 15.24	& 15.89	& 2.14	& 7.43	& 6.75	& 5.31	& 4.64	& --	& --	& --	& 5.19	& 4.55\\
NGC~3377	& 35.48	& 1.0	& 3.0	& 10.0	& 18.62	& 16.23	& 14.08	& 14.75	& 19.18	& 16.35	& 14.15	& 14.79	& 2.39	& 7.46	& 6.79	& 5.07	& 4.40	& 2.83	& 7.86	& 7.22	& 5.04	& 4.40\\
NGC~3458	& 11.48	& 1.9	& 3.0	& 10.0	& 21.22	& 19.39	& 16.89	& 17.61	& 20.91	& 19.05	& 17.02	& 17.84	& 1.83	& 7.28	& 6.56	& 5.49	& 4.76	& 1.86	& 6.92	& 6.03	& 5.07	& 4.18\\
NGC~4283	& 12.30	& 1.0	& 3.0	& 10.0	& 20.31	& 18.67	& 16.02	& 16.73	& 20.05	& 19.21	& 16.51	& 17.25	& 1.69	& 7.22	& 6.50	& 5.58	& 4.86	& 0.92	& 6.27	& 5.56	& 5.35	& 4.63\\
NGC~4346	& 19.50	& 2.1	& 3.0	& 10.0	& 20.67	& 18.56	& 16.03	& 16.74	& 19.59	& 17.64	& 15.42	& 16.12	& 2.13	& 7.56	& 6.85	& 5.47	& 4.75	& 1.96	& 7.13	& 6.41	& 5.20	& 4.48\\
NGC~4377	& 13.49	& 1.0	& 3.0	& 10.0	& 19.65	& 17.92	& 15.62	& 16.34	& --	& 19.01	& 17.06	& 17.69	& 1.74	& 6.96	& 6.23	& 5.26	& 4.53	& --	& --	& --	& --	& --\\
NGC~5500	& 15.14	& 1.5	& 3.0	& 10.0	& 21.62	& 19.50	& 17.47	& 18.18	& 20.92	& 19.12	& 17.64	& 18.29	& --	& --	& --	& 5.00	& 4.27	& --	& --	& --	& --	& --\\
PGC~05175	& 10.23	& 1.4	& 3.0	& 10.0	& 21.18	& 20.23	& 18.19	& 18.83	& --	& 19.97	& 18.98	& 19.38	& 0.94	& 6.02	& 5.36	& 5.12	& 4.46	& --	& --	& --	& 4.12	& 3.73\\
UGC~08876	& 8.51	& --	& 3.0	& 10.0	& --	& --	& 16.92	& 17.63	& 21.36	& 19.72	& 17.89	& 18.53	& --	& --	& --	& --	& --	& 1.66	& 6.68	& 6.01	& 5.11	& 4.43\\
\hline
\end{tabular}
}
\end{table}
\end{landscape}

\pagestyle{plain}

\begin{figure*}
  \centering
      \includegraphics{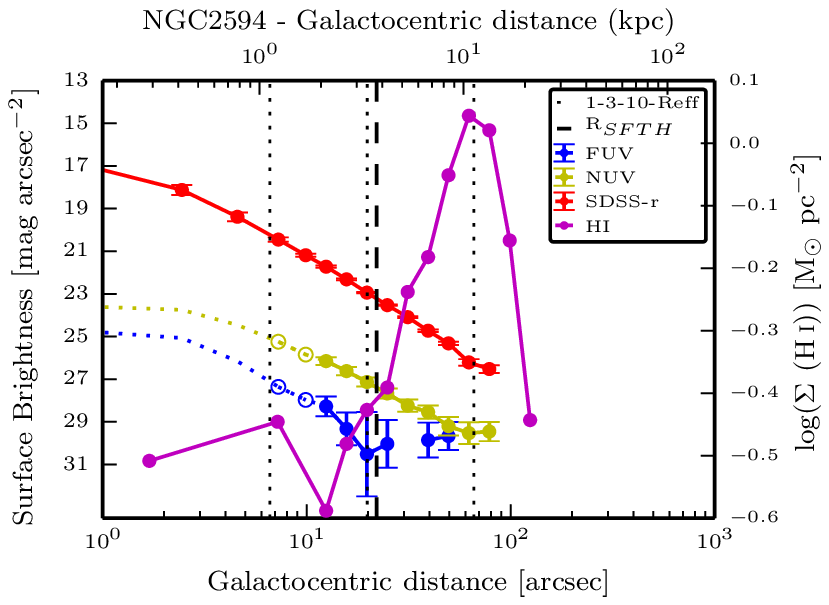}
      \hfill
      \includegraphics[scale=1.0]{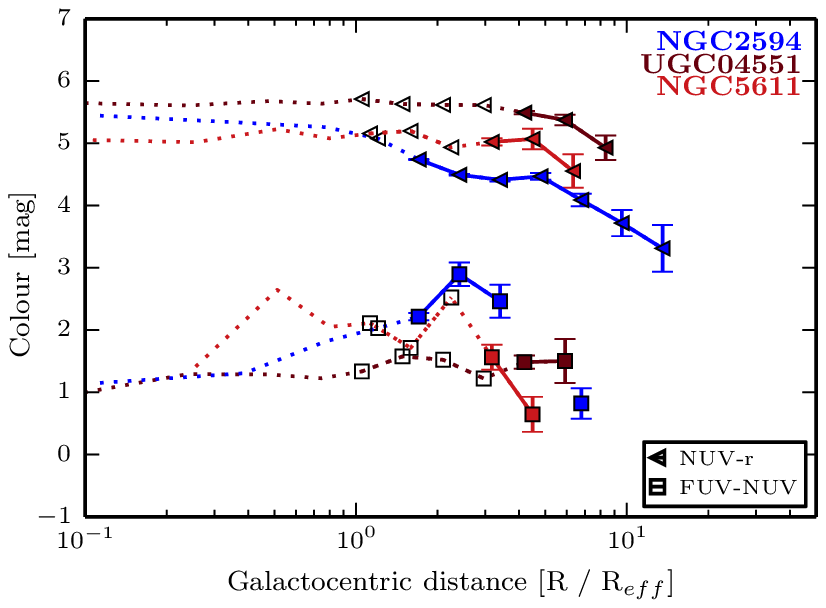}
\end{figure*}
 
\begin{figure*}
  \centering
  \includegraphics{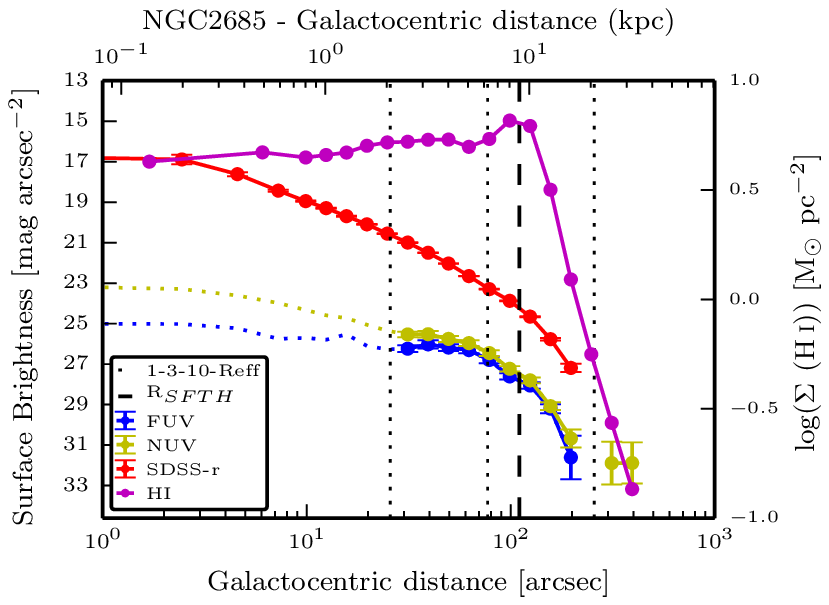}
  \hfill
  \includegraphics{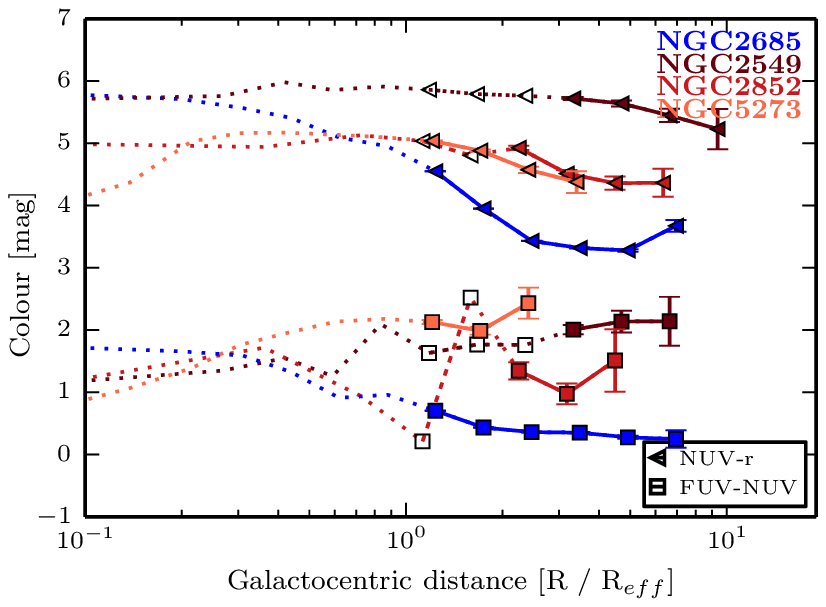}
\end{figure*}
\begin{figure*}
  \centering
  \includegraphics{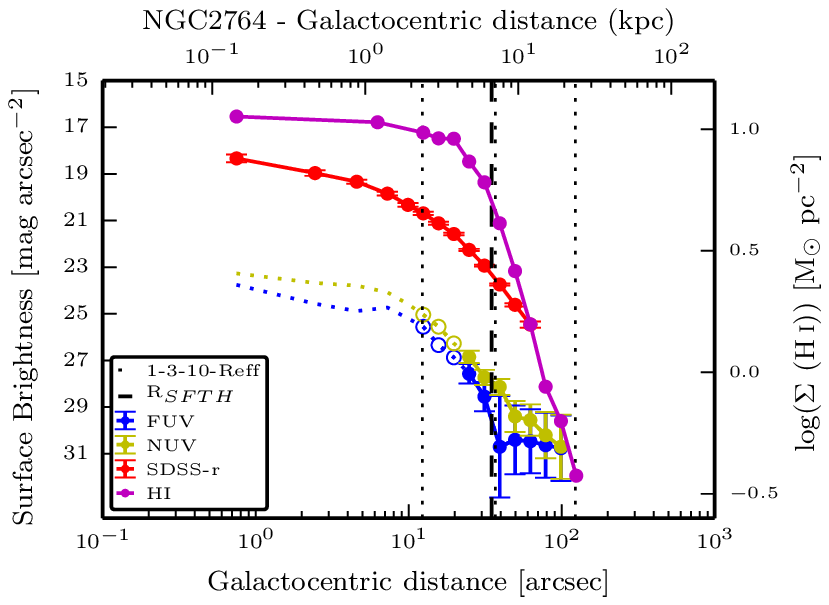}
  \hfill
  \includegraphics{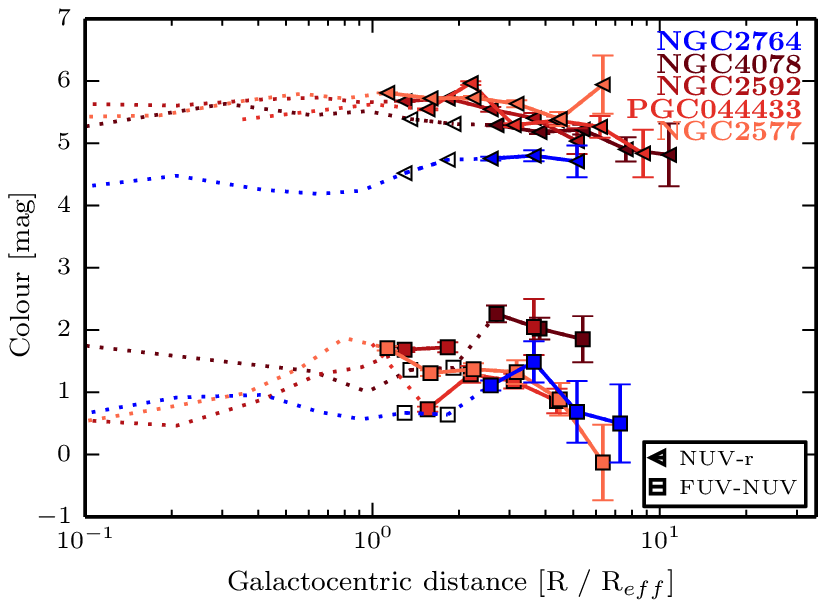}
  \caption{\textit{Left:} Azimuthally averaged surface brightness profiles of the FUV (blue), NUV (yellow), SDSS-r band (red) images, and surface density profile of the \hi \ image (magenta) for \hi-rich ETGs. The \hi \ surface density units are given on the right-y axis in magenta. The black-vertical dashed line show the star formation threshold (see Sec. \ref{sec:rad_pro}). The vertical dotted lines corresponds to 1, 3 and 10 R$_{eff}$, respectively. \textit{Right:} The colour profile of the \hi-rich galaxy (blue) and its control galaxies (red shades). The plots are shown in one row for one \hi-rich galaxy. Open symbols indicate annuli below the secure radius, where the PSF contamination is large (Sec. \ref{sec:psf_effect}). The error bars represent the uncertainty in determining the background.}
\label{fig:appendix}
\end{figure*}

\begin{figure*}
  \centering
  \includegraphics{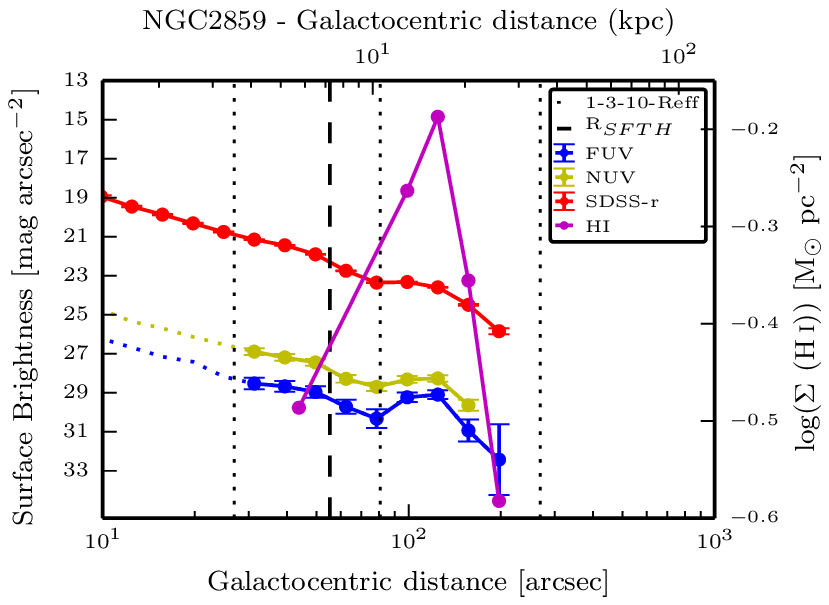}
  \hfill
  \includegraphics{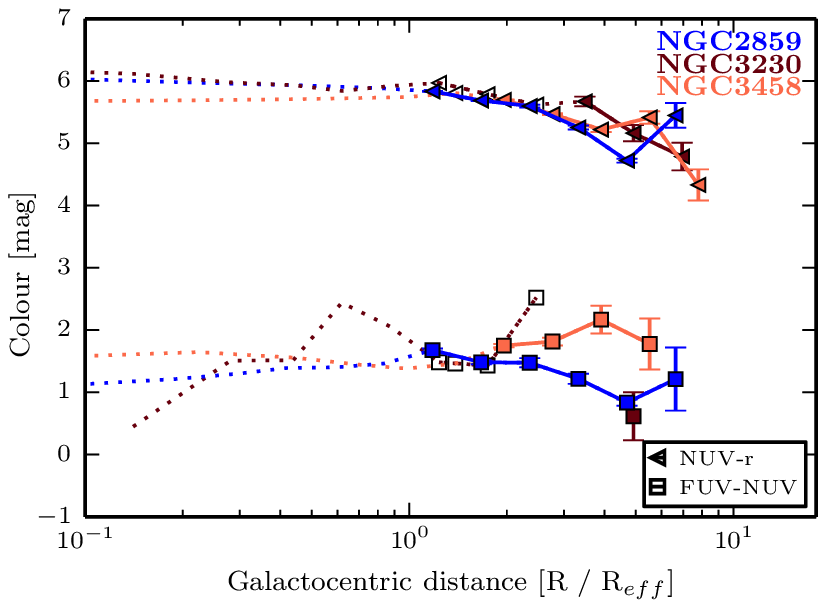}
\end{figure*}
\begin{figure*}
  \centering
  \includegraphics{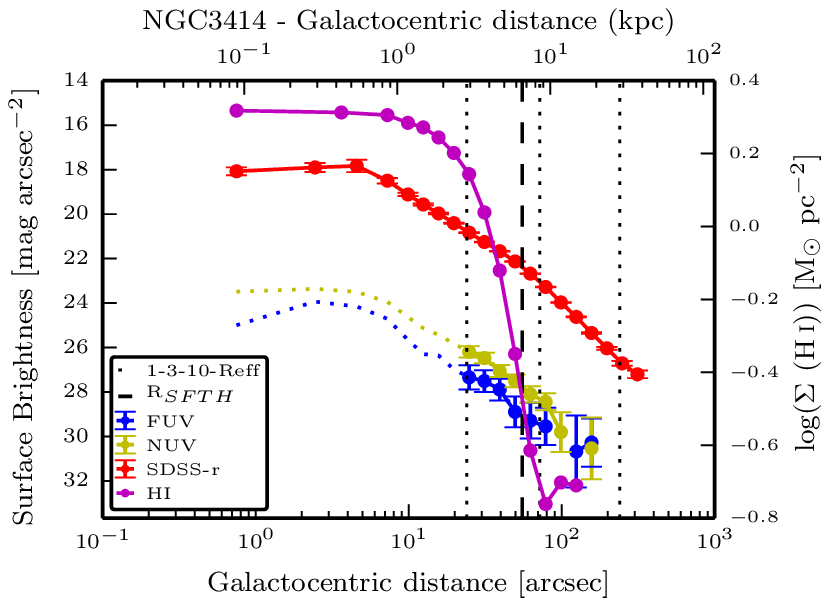}
  \hfill
  \includegraphics{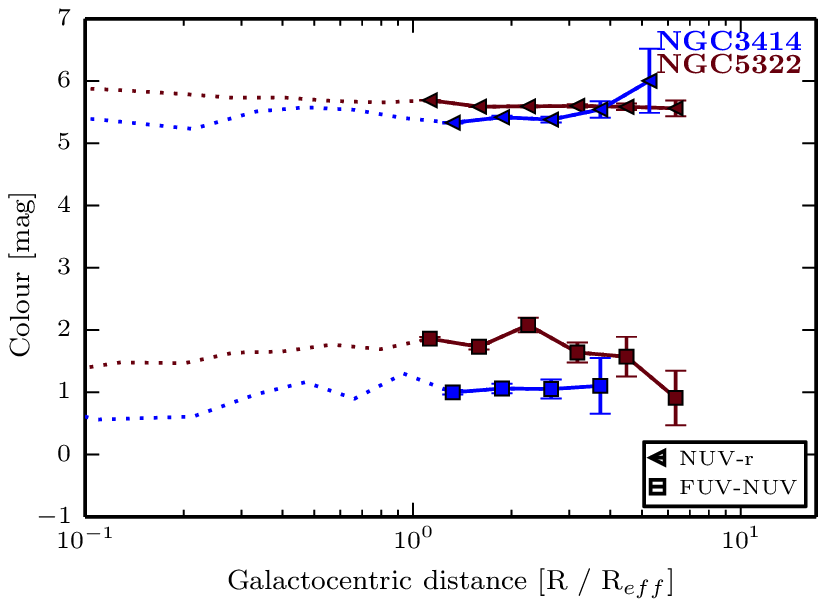}
\end{figure*}
\begin{figure*}
  \centering
  \includegraphics{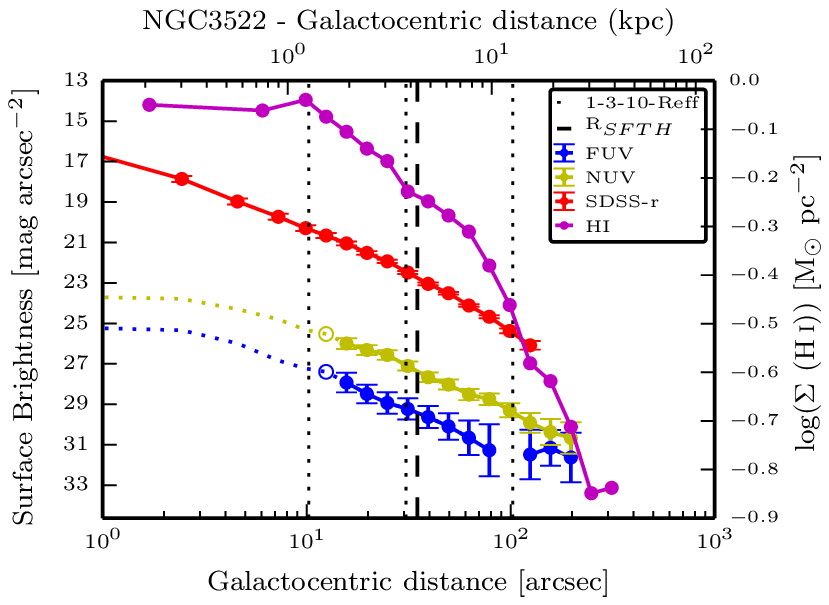}
  \hfill
  \includegraphics{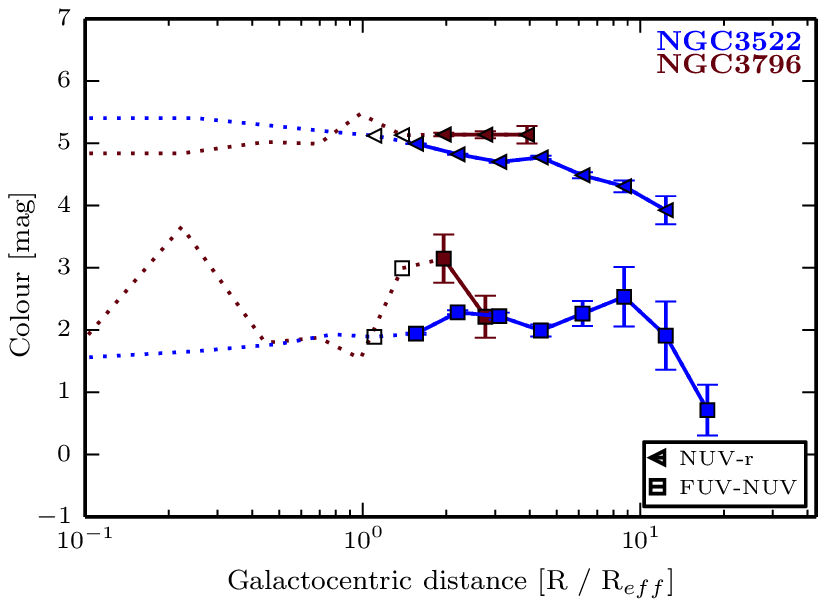}
  \caption{Same as Figure \ref{fig:appendix}}
\end{figure*}

\begin{figure*}
  \centering
  \includegraphics{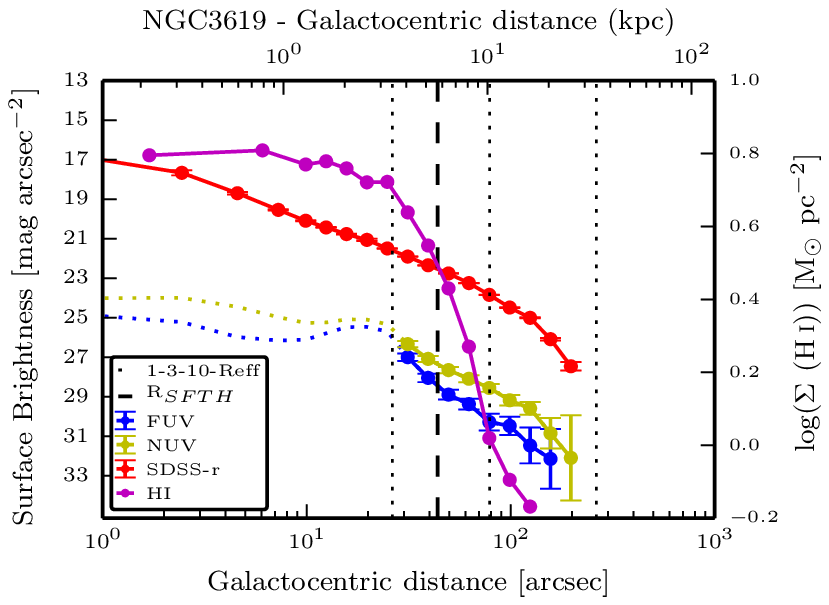}
  \hfill
  \includegraphics{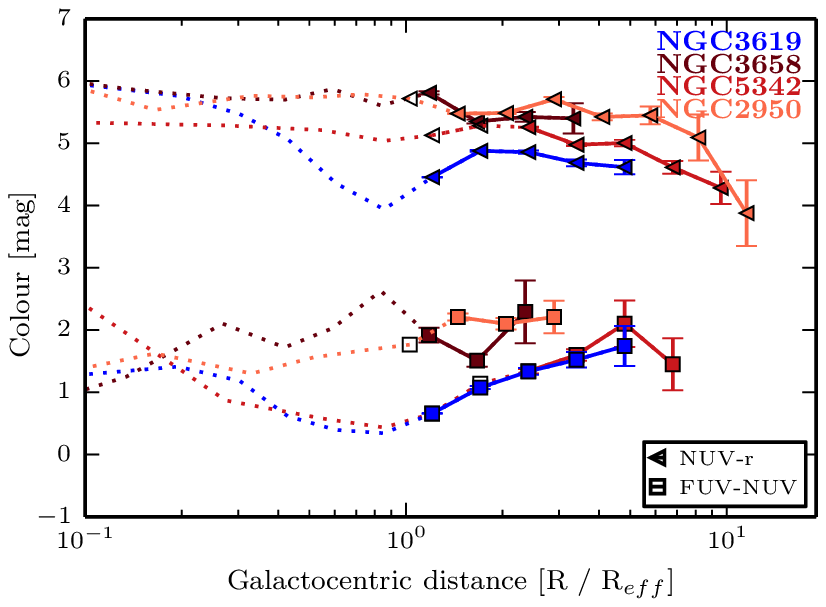}
\end{figure*}
\begin{figure*}
  \centering
  \includegraphics{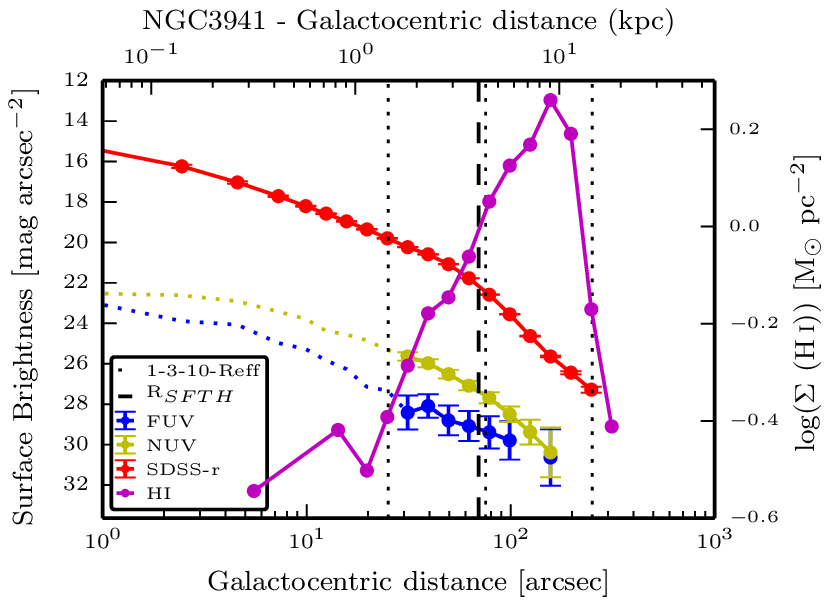}
  \hfill
  \includegraphics{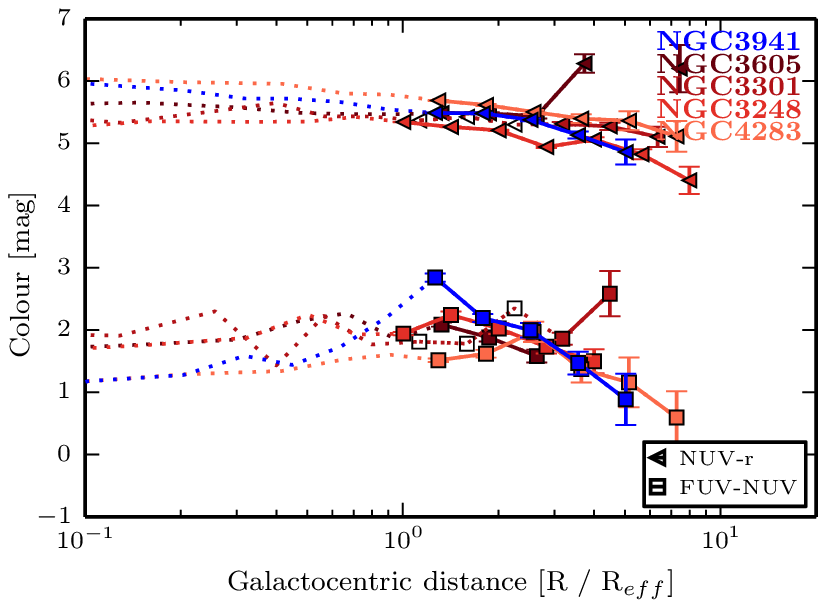}
\end{figure*}
\begin{figure*}
  \centering
  \includegraphics{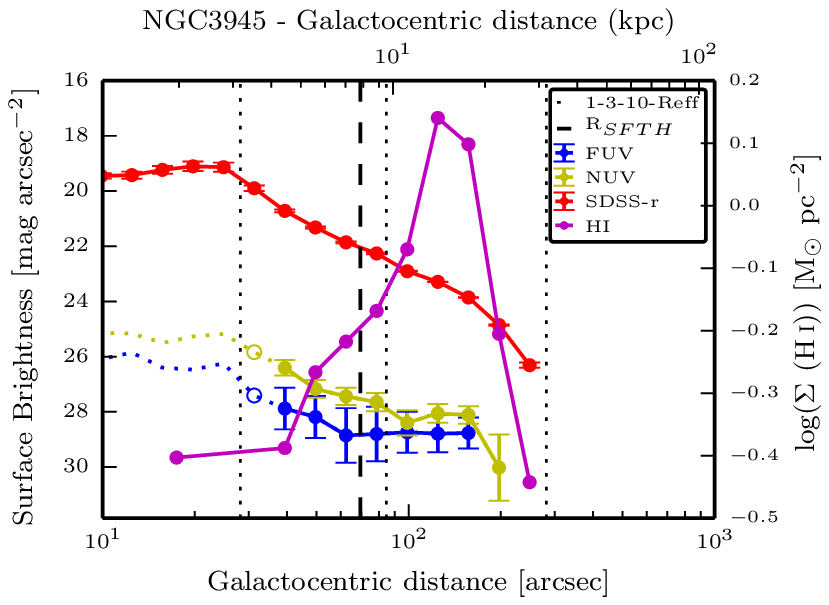}
  \hfill
  \includegraphics{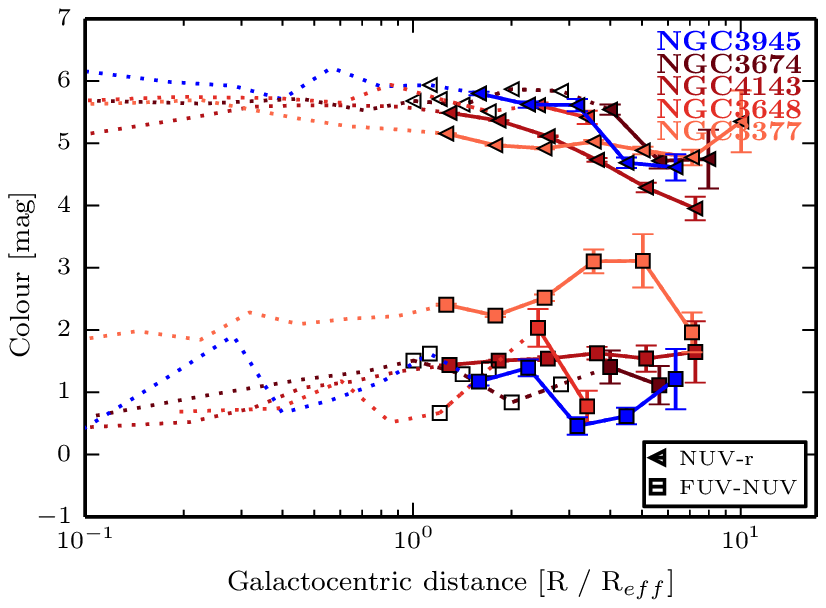}
  \caption{Same as Figure \ref{fig:appendix}}
\end{figure*}

\begin{figure*}
  \centering
  \includegraphics{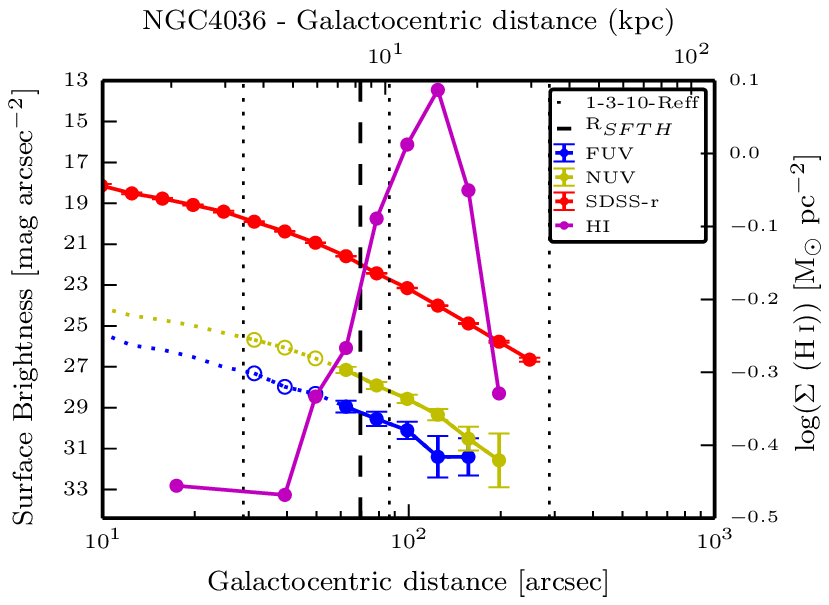}
  \hfill
  \includegraphics{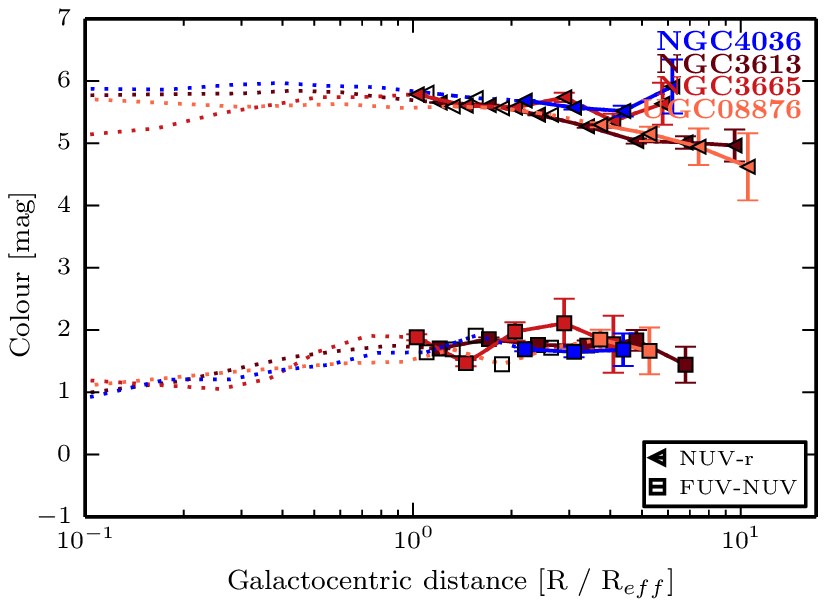}
\end{figure*}
\begin{figure*}
  \centering
  \includegraphics{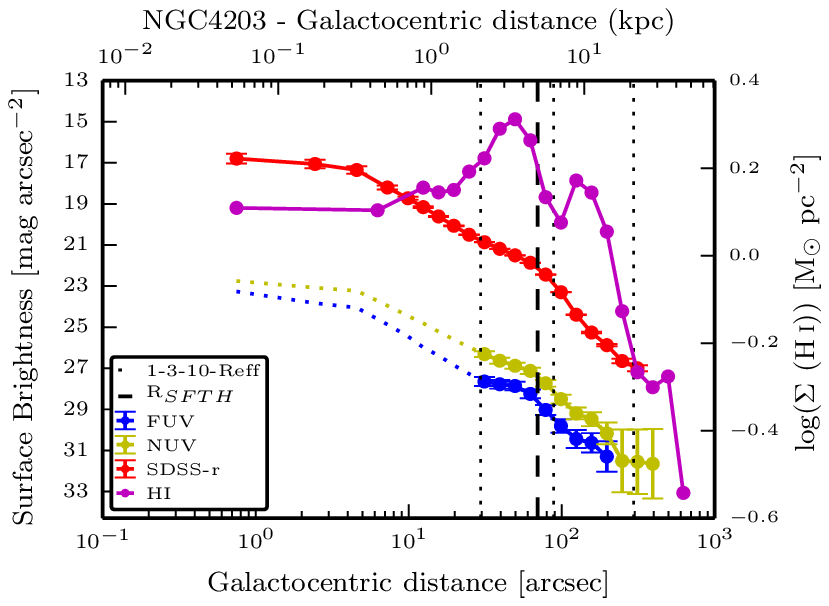}
  \hfill
  \includegraphics{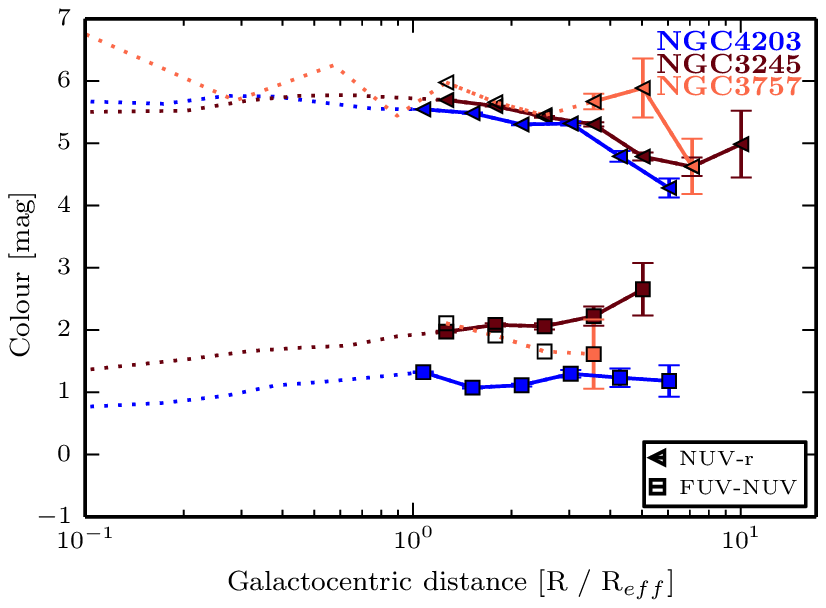}
\end{figure*}
\begin{figure*}
  \centering
  \includegraphics{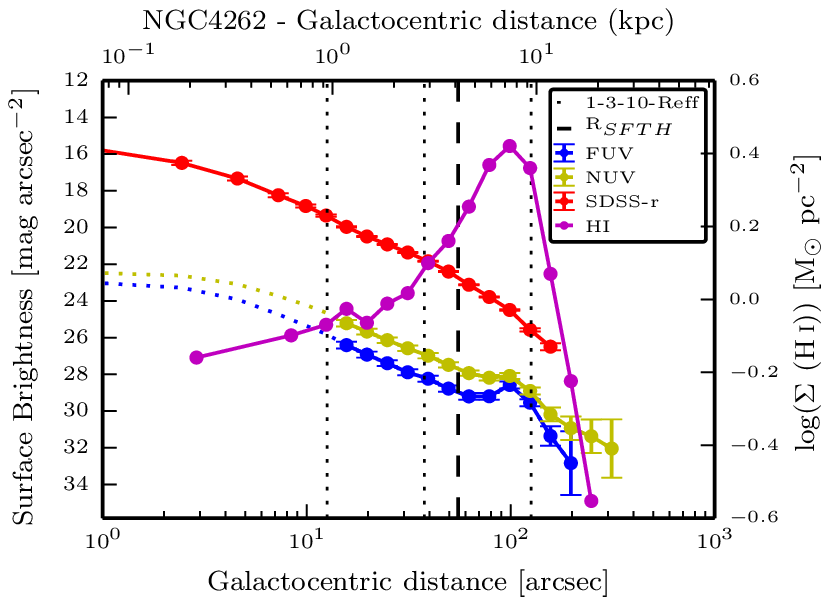}
  \hfill
  \includegraphics{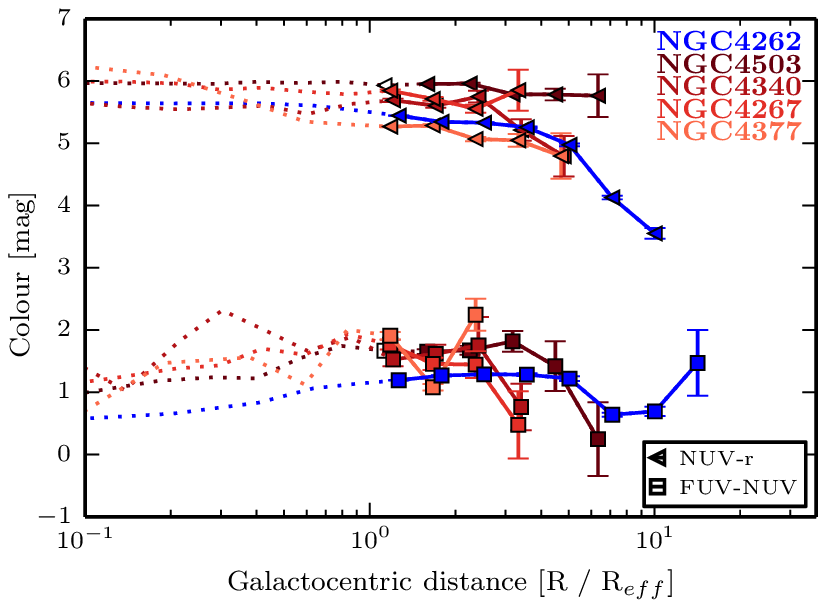}
  \caption{Same as Figure \ref{fig:appendix}}
\end{figure*}

\begin{figure*}
  \centering
  \includegraphics{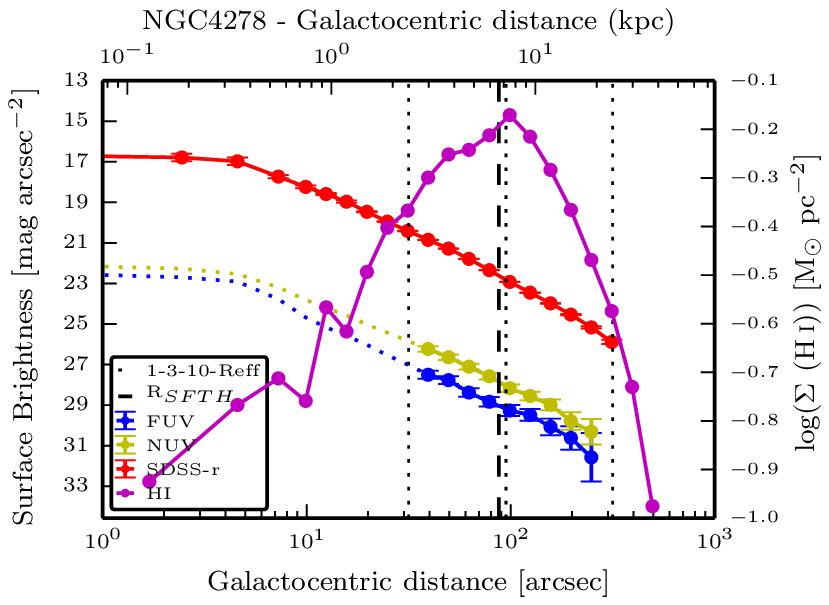}
  \hfill
  \includegraphics{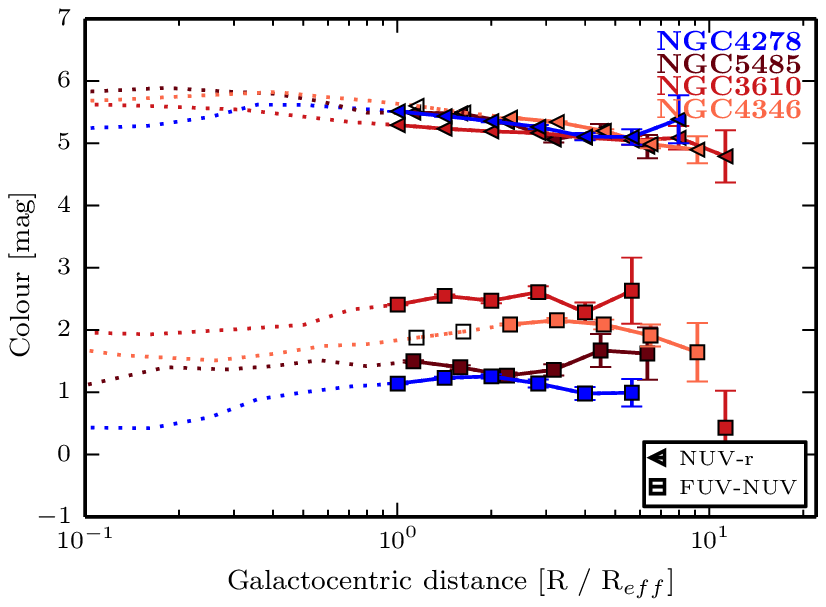}
\end{figure*}
\begin{figure*}
  \centering
  \includegraphics{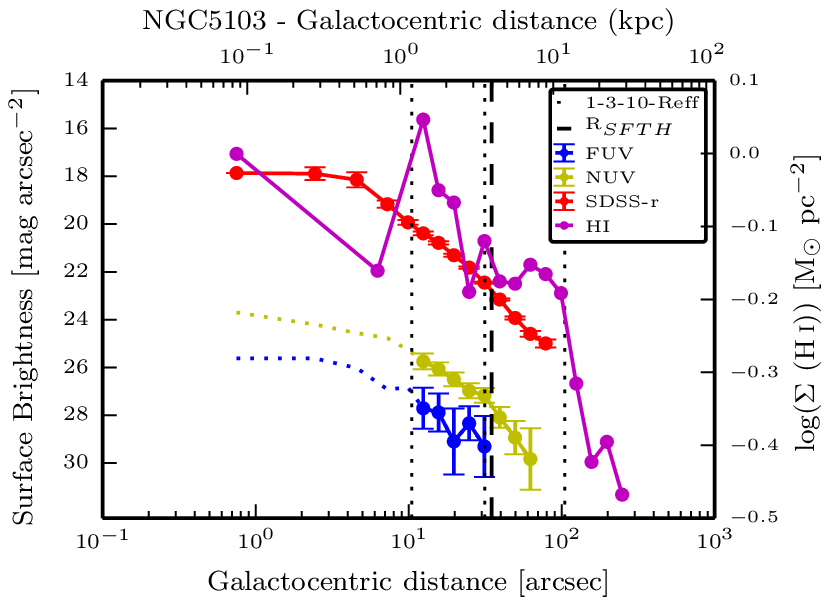}
  \hfill
  \includegraphics{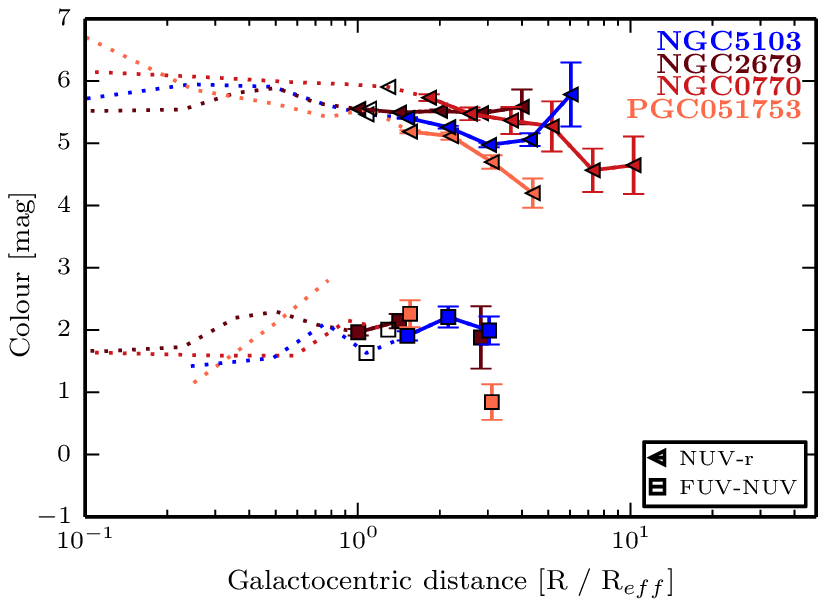}
\end{figure*}
\begin{figure*}
  \centering
  \includegraphics{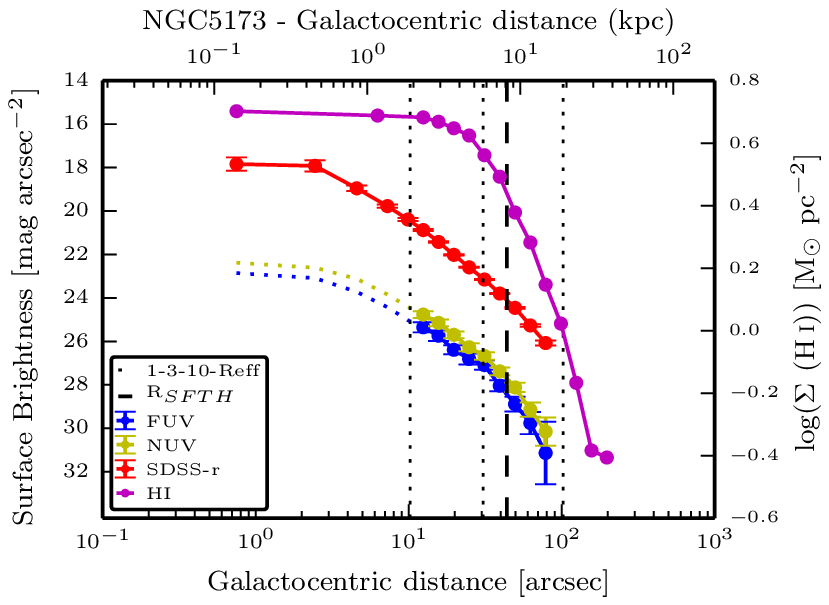}
  \hfill
  \includegraphics{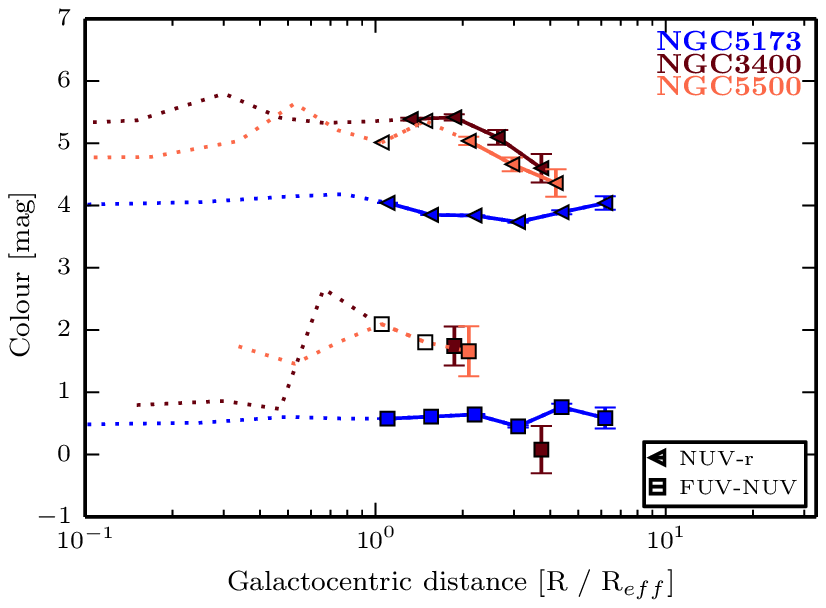}
  \caption{Same as Figure \ref{fig:appendix}}
\end{figure*}

\begin{figure*}
  \centering
  \includegraphics{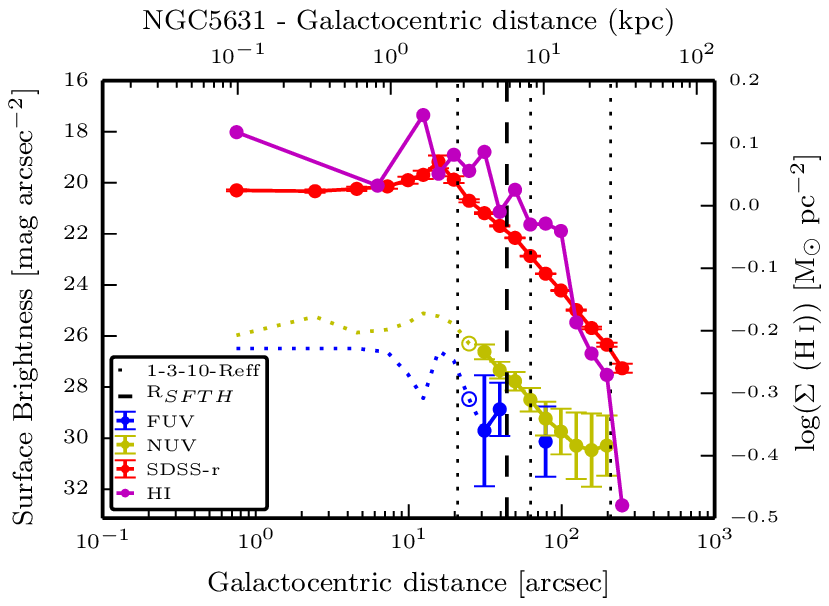}
  \hfill
  \includegraphics{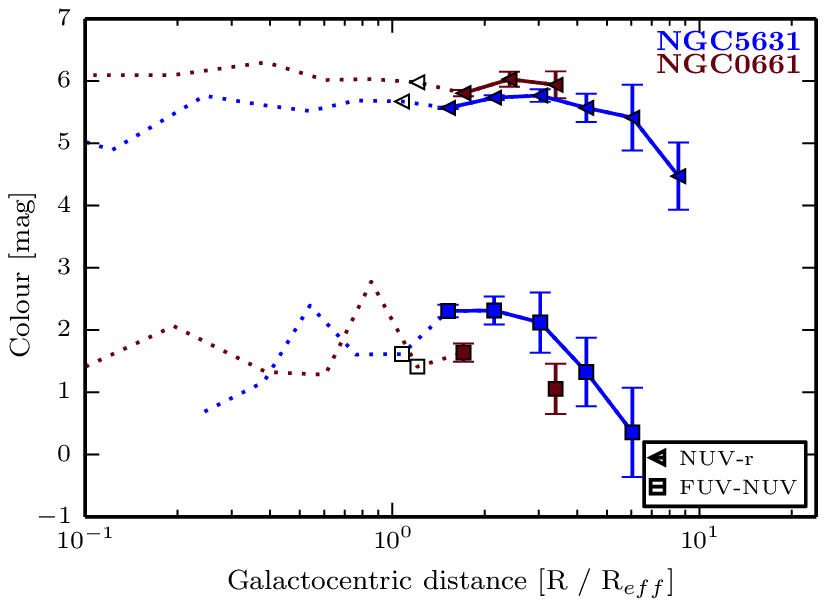}
\end{figure*}
\begin{figure*}
  \centering
  \includegraphics{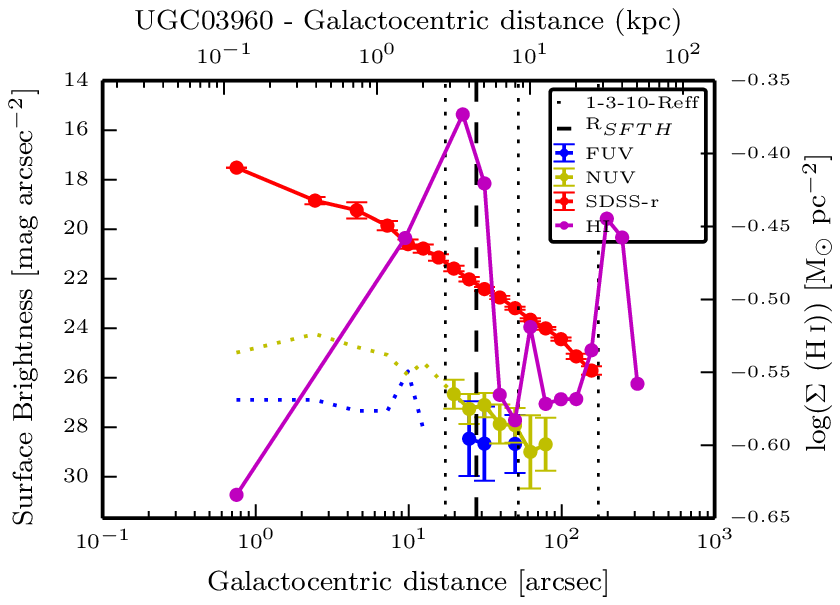}
  \hfill
  \includegraphics{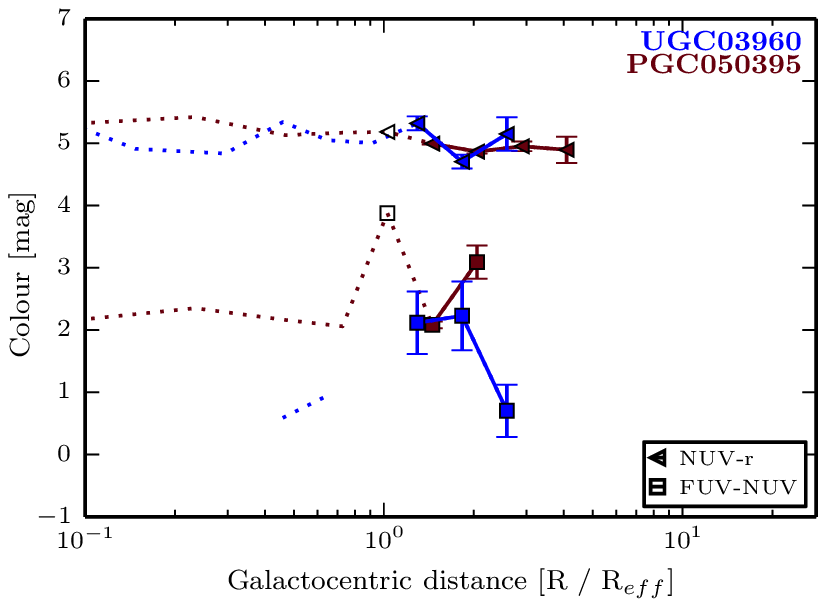}
\end{figure*}
\begin{figure*}
  \centering
  \includegraphics{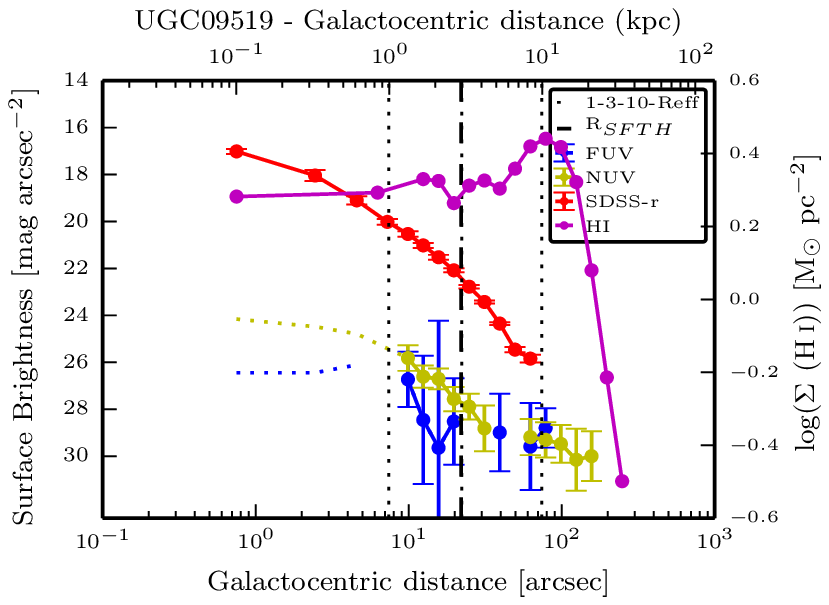}
  \hfill
  \includegraphics{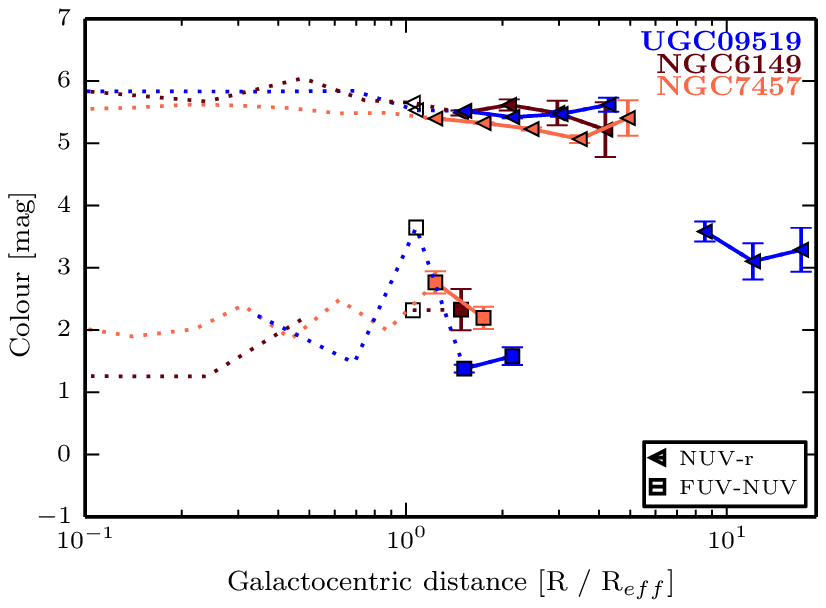}
  \caption{Same as Figure \ref{fig:appendix}}
\end{figure*}

\clearpage
\pagestyle{plain}

\begin{figure*}
  \centering
  \includegraphics{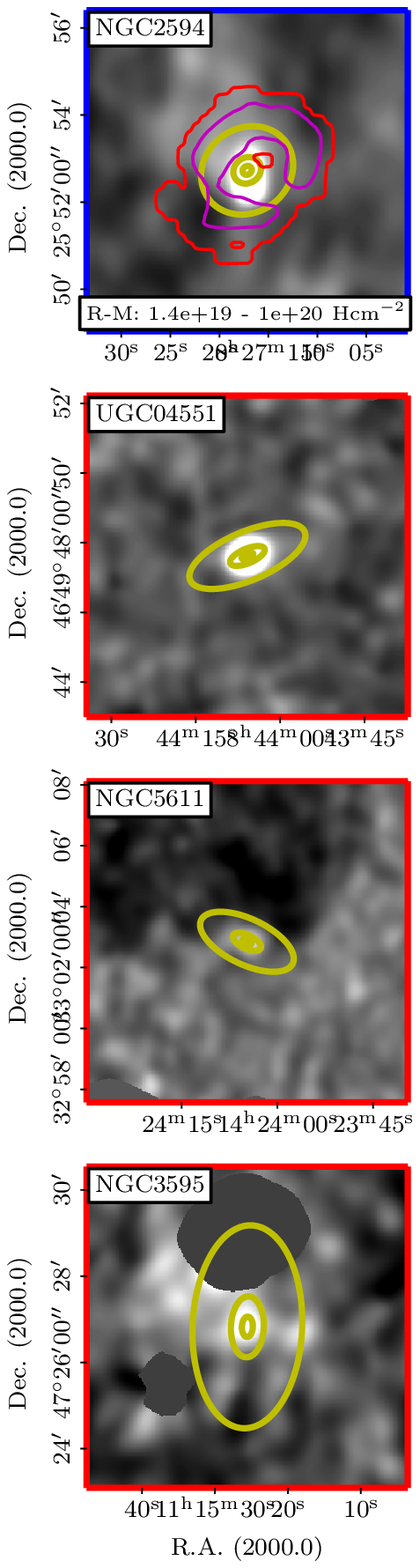}
  \includegraphics{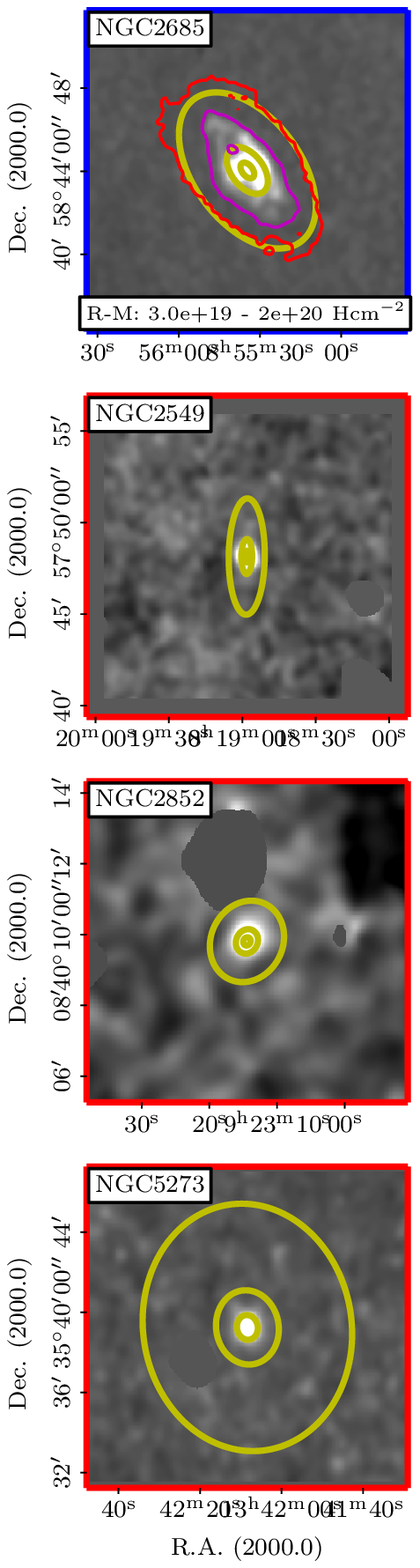}
  \includegraphics{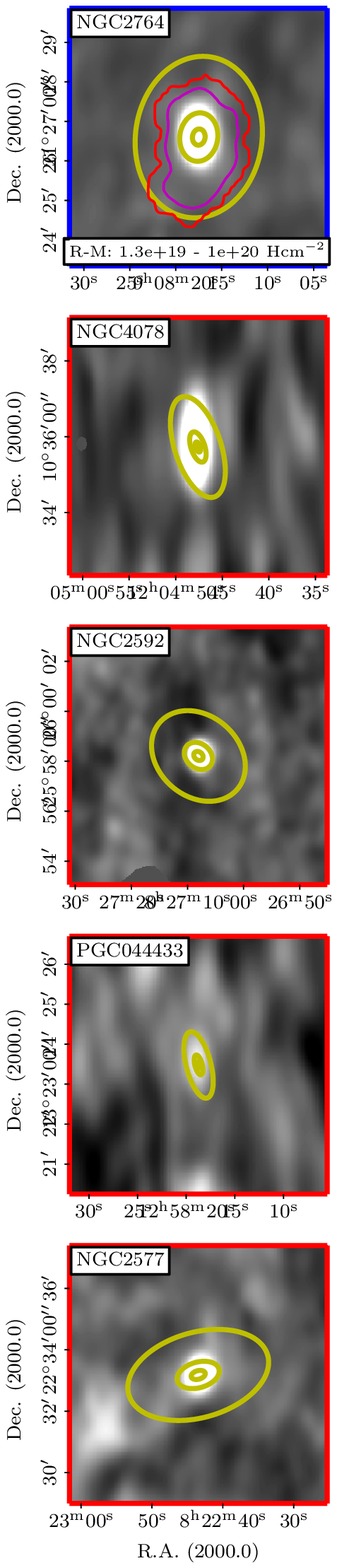}
  \caption{\textit{Blue boxes:} FUV images of the \hi-rich galaxies, smoothed to the \hi \ resolution, together with \hi-column density contours (in red and magenta). \textit{Red boxes:} Smoothed FUV images of the selected \hi-poor control galaxies for the \hi-rich galaxies are shown on the top in blue box. The two apertures used in this study are indicated with 3 yellow ellipses at 1-3-10 $R_\mathrm{eff}$ (see Sec. \ref{sec:rad_pro} for PA and ellipticity parameters). All images show a field of size 150 kpc $\times$ 150 kpc.}
\label{fig:appendix3}
\end{figure*}

\begin{figure*}
  \centering
  \includegraphics{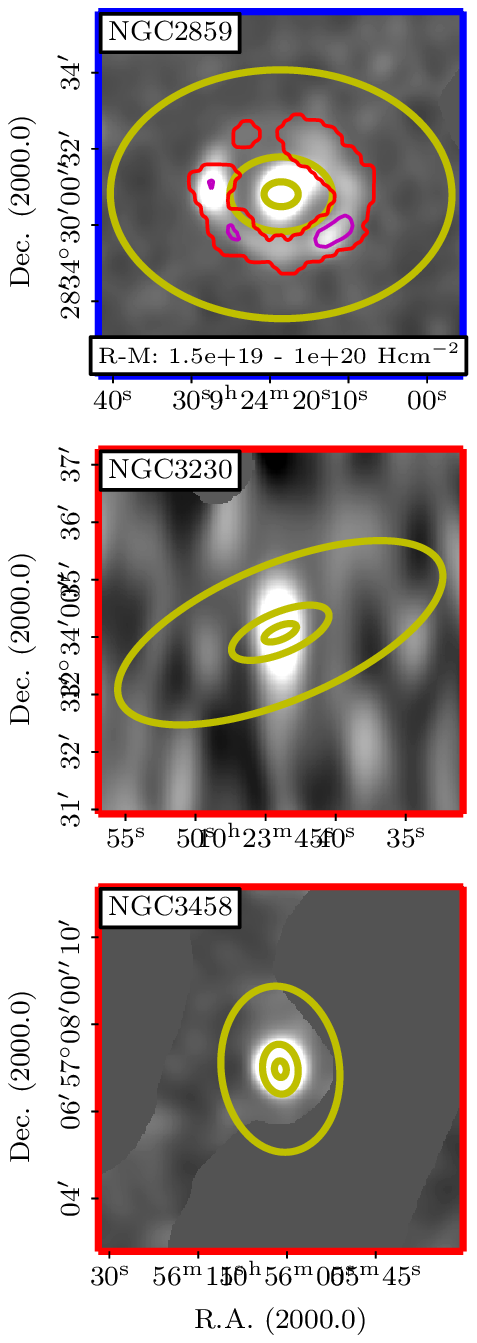}
  \includegraphics{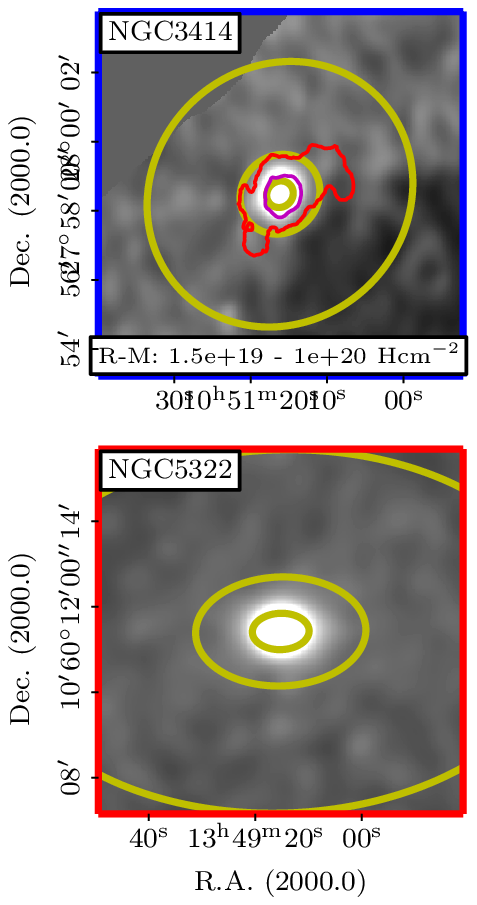}
  \includegraphics{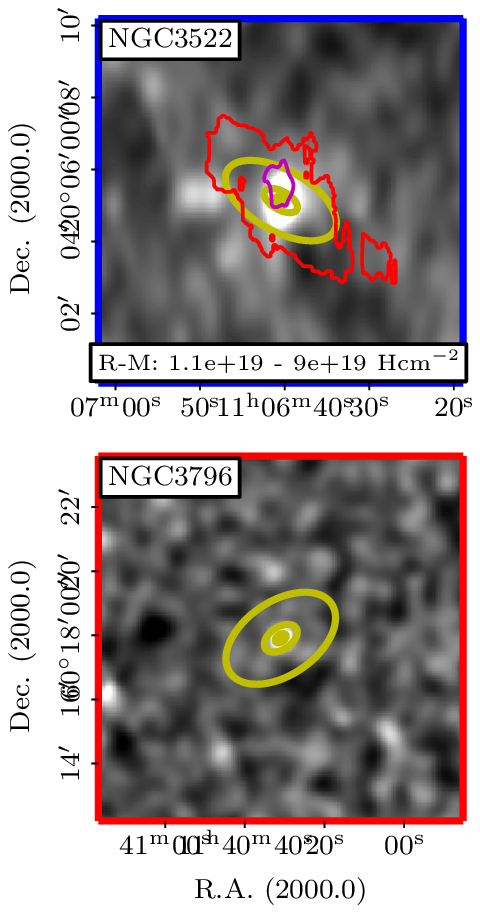}
  \caption{Same as Figure \ref{fig:appendix3}} 
\end{figure*}

\begin{figure*}
  \centering
  \includegraphics{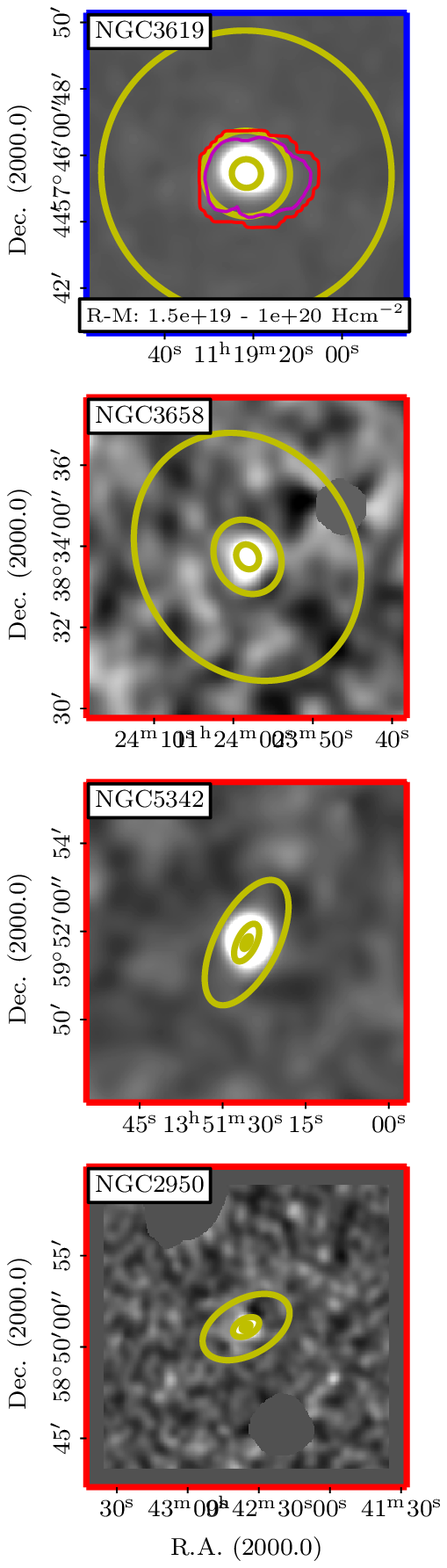}
  \includegraphics{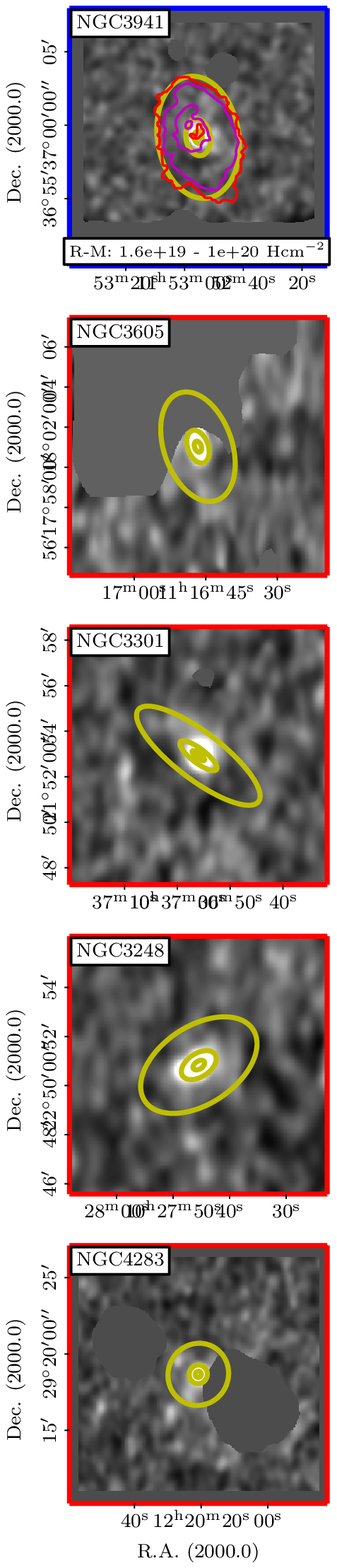}
  \includegraphics{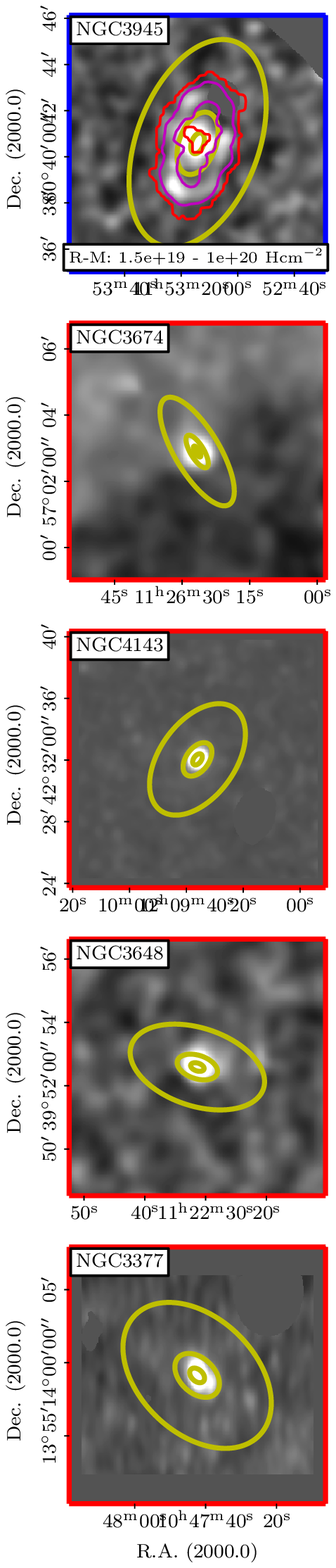}
  \caption{Same as Figure \ref{fig:appendix3}} 
\end{figure*}

\begin{figure*}
  \centering
  \includegraphics{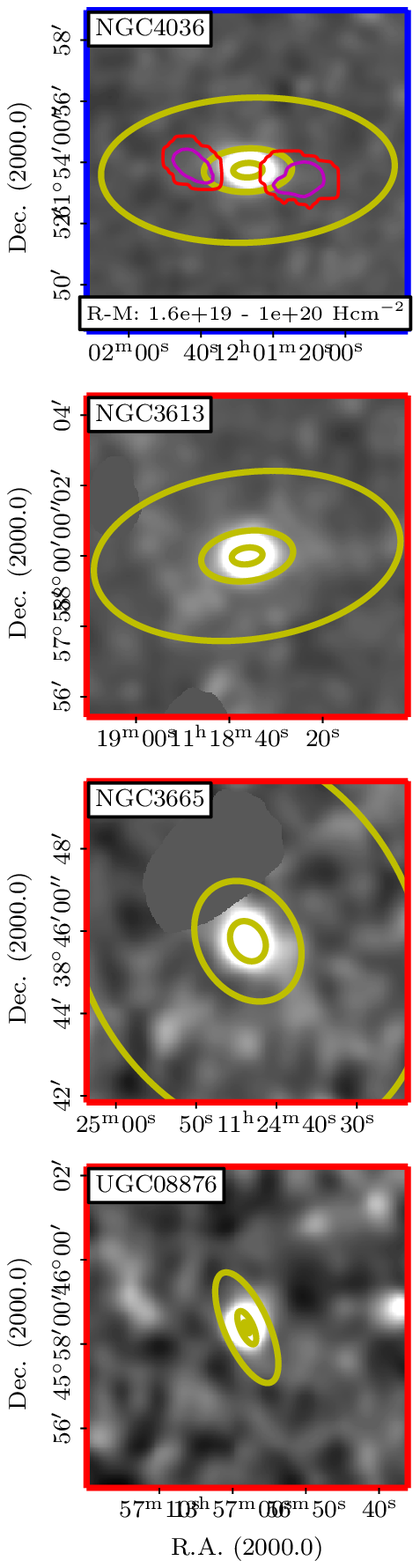}
  \includegraphics{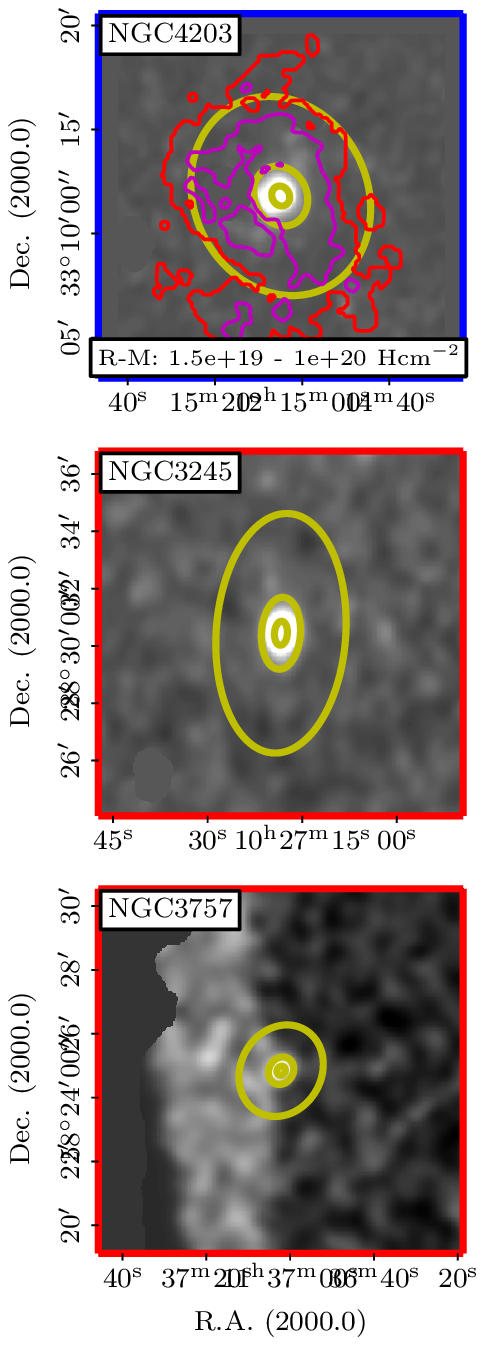}
  \includegraphics{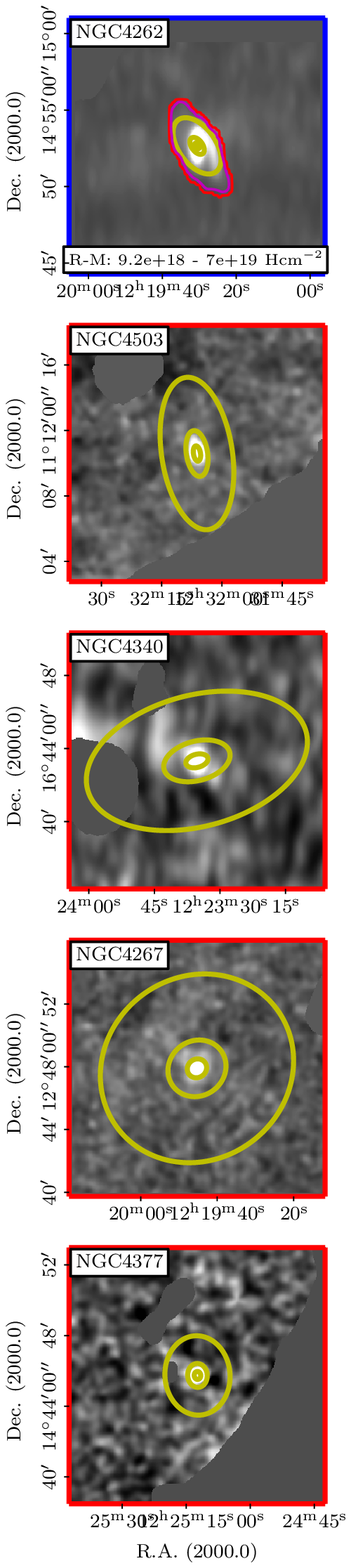}
  \caption{Same as Figure \ref{fig:appendix3}}
\end{figure*}

\begin{figure*}
  \centering
  \includegraphics{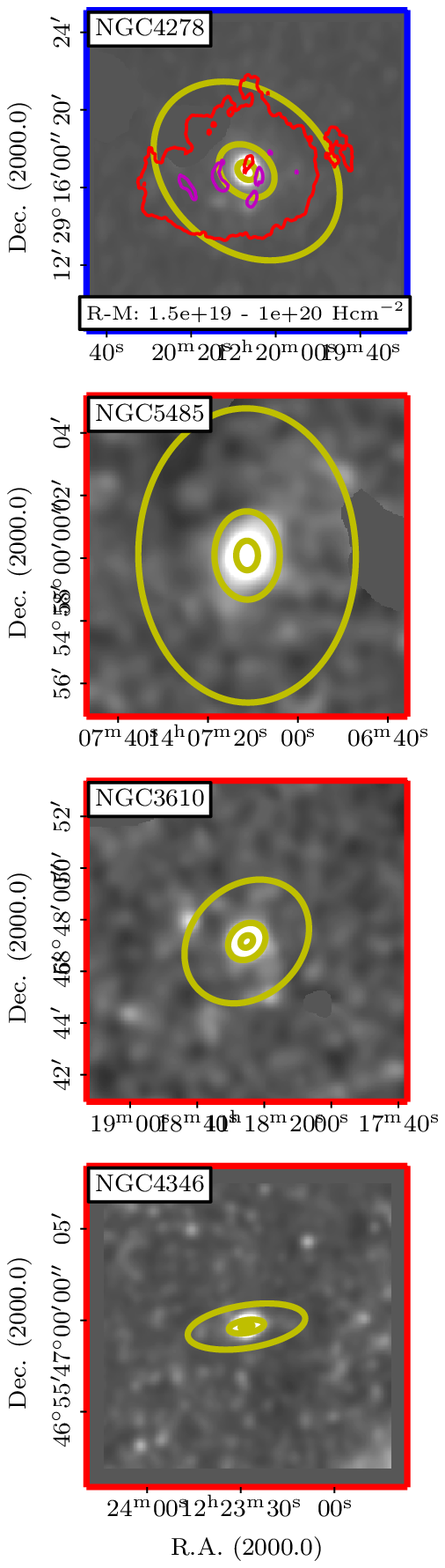}
  \includegraphics{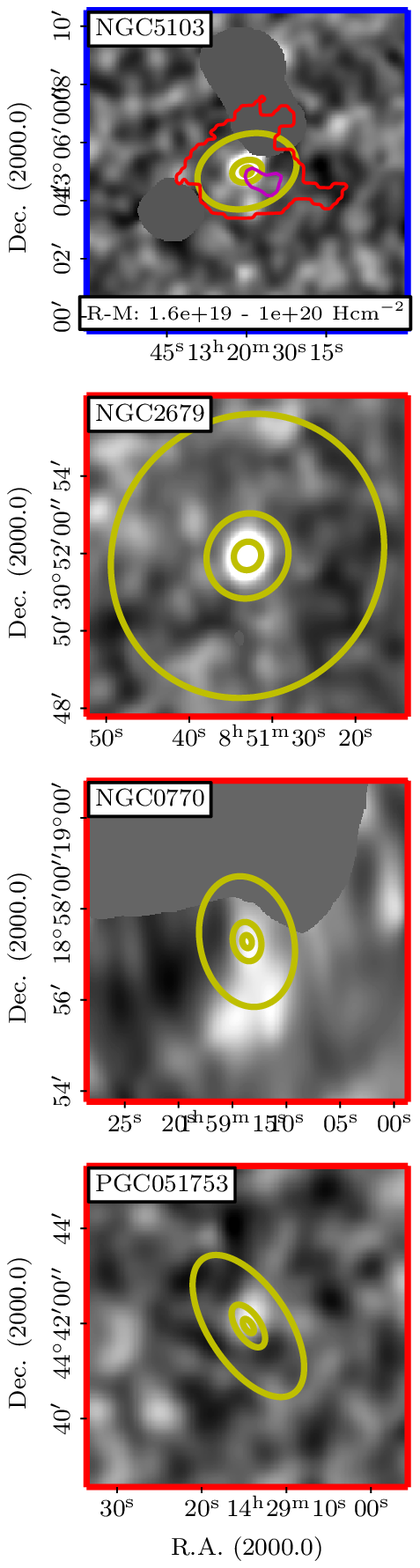}
  \includegraphics{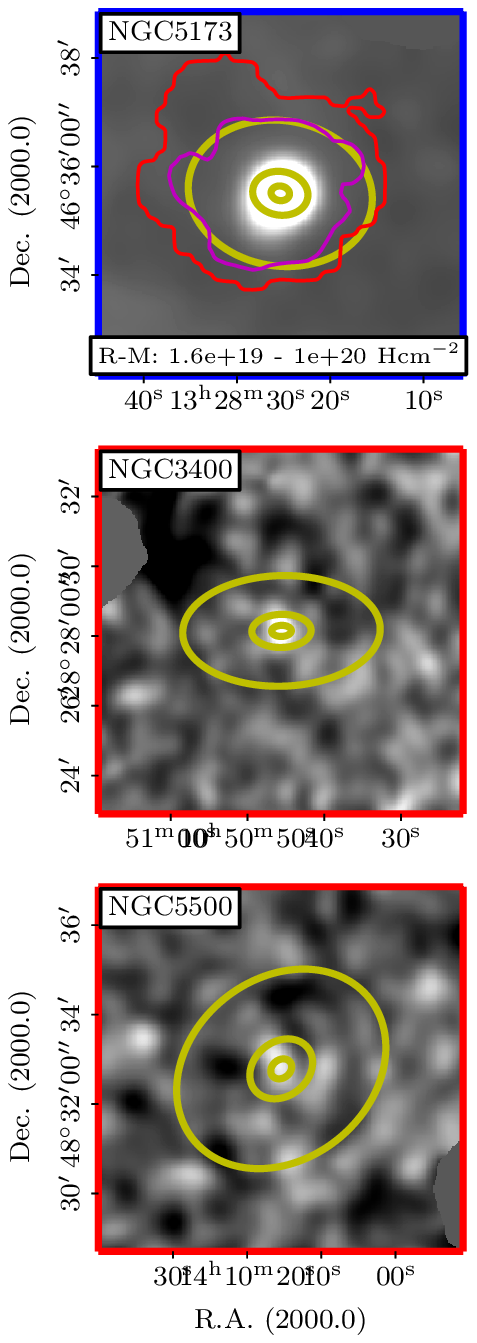}
  \caption{Same as Figure \ref{fig:appendix3}}
\end{figure*}

\begin{figure*}
  \centering
  \includegraphics{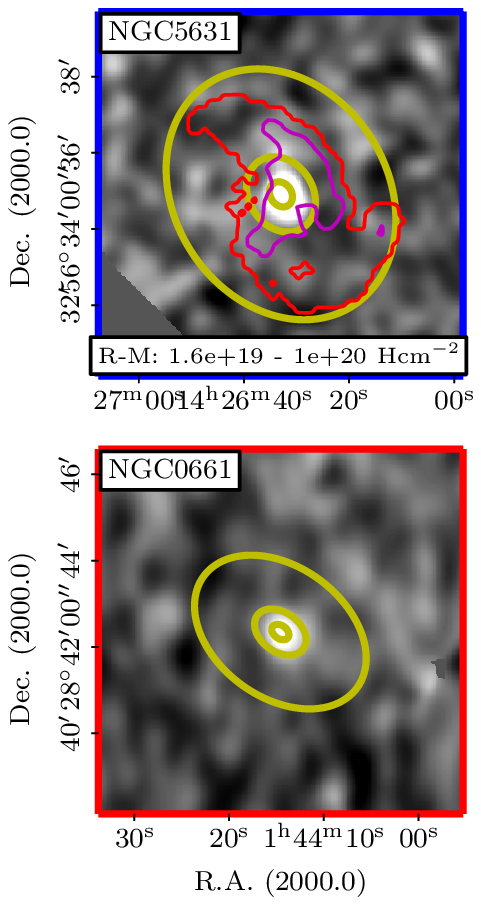}
  \includegraphics{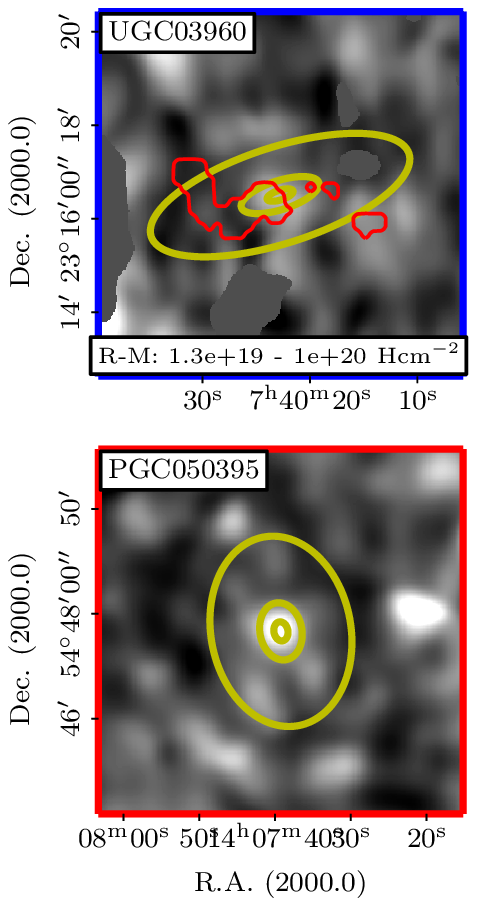}
  \includegraphics{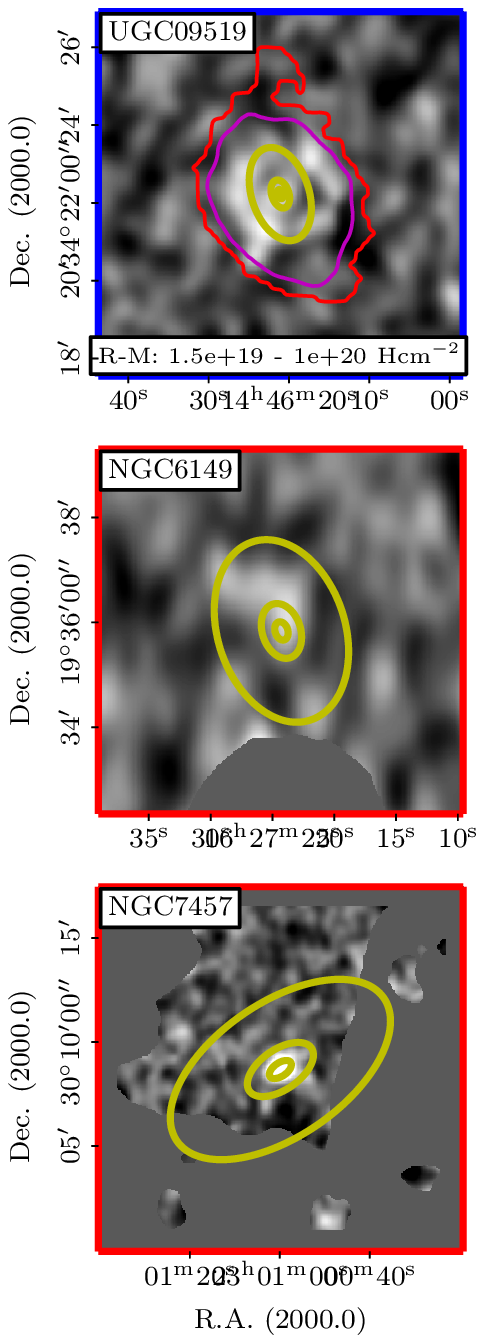}
  \caption{Same as Figure \ref{fig:appendix3}}
\end{figure*}

\end{document}